\documentclass[12pt]{article}

\usepackage{chemformula}
\usepackage[T1]{fontenc}
\usepackage[top=5em,bottom=5em,left=5em,right=5em]{geometry}
\usepackage{amsmath,amssymb}
\usepackage{siunitx,bm}
\usepackage{pdfpages}
\usepackage{anyfontsize}
\usepackage{epsdice}
\usepackage{xfrac}
\usepackage{multirow}
\usepackage{ulem}
\usepackage{authblk}
\usepackage{xr}
\usepackage[backend=biber,style=ieee,url=true,doi=true]{biblatex}
\usepackage{hyperref}
\definecolor{linkcolor}{rgb}{0.5,0.1,0.1}
\definecolor{urlcolor} {rgb}{0.1,0.1,0.5}
\definecolor{citecolor}{rgb}{0.1,0.5,0.1}
\hypersetup{
 colorlinks=true,
 linkcolor=linkcolor,
 urlcolor=urlcolor,
 citecolor=citecolor,
 pdfauthor={},
 pdftitle={},
 pdfkeywords={},
 pdfsubject={},
 pdfcreator={}, 
 pdflang={English}}

\newcommand*{\tn}{\ensuremath{\epsdice{1}}}
\newcommand*{\rods}{\ensuremath{\epsdice[black]{5}}}
\newcommand*{\figref}[1]{Figure \ref{#1}}

\renewcommand*{\eqref}[1]{Eq.~(\ref{#1})}
\newcommand*{\inl}[1]{\,{#1}\,}
\newcommand*{\eq}{\inl=}
\newcommand*{\deq}{\inl{:=}}

\newcommand*{\mat}[1]{\bm{#1}}

\definecolor{Gray}{gray}{0.9}
\newcolumntype{a}{>{\columncolor{Gray}}c}
\renewcommand{\emph}[1]{\textit{#1}}

\newcommand{\delete}[1]{}
\newcommand{\add}[1]{#1}
\newcommand{\replace}[2]{#2}

\title{Termination-Driven Control over BIC Q-Factors and Frequencies in Plasmonic Double Net Metamaterials}

\author[1]{C\'edric Schumacher}
\author[1]{Bilel Abdennadher}
\author[1,2]{Ullrich Steiner}
\author[1,2,*]{Matthias Saba}
\affil[1]{Adolphe Merkle Institute, University of Fribourg, Chemin des Verdiers 4, 1700 Fribourg, Switzerland}
\affil[2]{NCCR Bioinspired Materials, University of Fribourg, 1700 Fribourg, Switzerland}
\affil[*]{\href{mailto:matthias.saba@unifr.ch}{matthias.saba@unifr.ch}}

\begin{document}
\maketitle

\begin{abstract}
Interlaced metallic wire meshes are 3D metamaterials consisting of two intertwined metallic networks. These plasmonic double nets give rise to otherwise unobserved longitudinal, weakly dispersive and broadband \emph{electron acoustic modes} from the effective plasma frequency of the double net down to arbitrarily low frequencies.
These modes have recently been shown to generate confined slab modes with extremely long lifetimes (high quality factors), so-called \emph{quasi-bound states in the continuum}.
This work reveals the central role of the double net termination in determining the mode's resonant frequency and quality factor.
We compare two limiting cases, a \emph{tennis net} termination recently studied experimentally by others and a protruding column array with a much lower quality factor, as demonstrated by microwave transmission experiments and full-wave simulations.
Our work thus vividly demonstrates the failure of a homogenisation approach to explain and quantify the physics of terminated plasmonic network materials.
We introduce a new approach, in which additional evanescent bulk states are included in the scattering problem, yielding a qualitative understanding of the slab's optical response.
The resulting engineering principles pave the way for the design and exploitation of these materials for applications such as coherent light generation.
\end{abstract}


\section*{Introduction}

The optical response of a metal can be understood from a Drude electron model, which predicts an absence of propagating modes below the metal's plasma frequency (proportional to the square root of the free electron density) \cite{Saleh2019}.
Metallic single meshes behave similarly to dilute metals with reduced electron density, where low-loss electromagnetic modes exist only above a reduced \emph{effective} plasma frequency, featuring the two usual transverse bands and a dispersive longitudinal band \cite{Sievenpiper1996}.
As in bulk metals, only evanescent modes exist below the effective plasma frequency in a wire mesh, resulting in a mirror-like behaviour.
Homogenisation models considering the interconnected topology of the network qualitatively predict its response and explain the underlying physical mechanisms \cite{Silve2005,Deme2008,dolan2016}.

A pair of intertwined metallic networks -- \emph{plasmonic double-nets} (PDNs) -- additionally exhibit a continuous longitudinal photonic band between zero frequency and the effective plasma frequency of the double-net \cite{Wang2023}.
Interestingly, these double-net structures homogenise in the low-frequency limit to a \emph{non-Maxwellian} effective medium \cite{Shin2007a}; this is a metamaterial that cannot be described by a local effective permittivity and permeability alone.
PDNs which interchange networks upon a primitive lattice translation emanate from a finite wave vector at the edge of the Brillouin zone at zero frequency \cite{chen2018}.
This leads to a unique low-frequency behaviour that can be engineered by the choice of PDN \cite{sakhno2021}.

Slab-terminated PDN materials excited by a plane wave exhibit an exotic light tunnelling anomaly \cite{Latioui2017}, reminiscent of a Fabry-P\'erot resonator with extremely high finesse.
A non-Maxwellian homogenisation of the PDN bulk by a warm decoupled double plasma model explains the coupling between the vacuum field and the longitudinal \emph{electron acoustic waves} (EAWs) within the PDN slab \cite{Wang2023}.
This leads to bound states in the continuum (BIC) \cite{Hsu2016} with infinite lifetime (or quality factor) at the centre of the 2D surface Brillouin zone of the slab (corresponding to a normal incidence radiation condition).
For non-normal radiation, quasi-BIC bands with finite but extremely long lifetimes emanate from the BICs at zero wave number.
As for most quasi-BIC states \cite{koshelev2018}, the quality factor scales inversely quadratically with the wavenumber.

\begin{figure}[htbp]
\includegraphics[width=\textwidth]{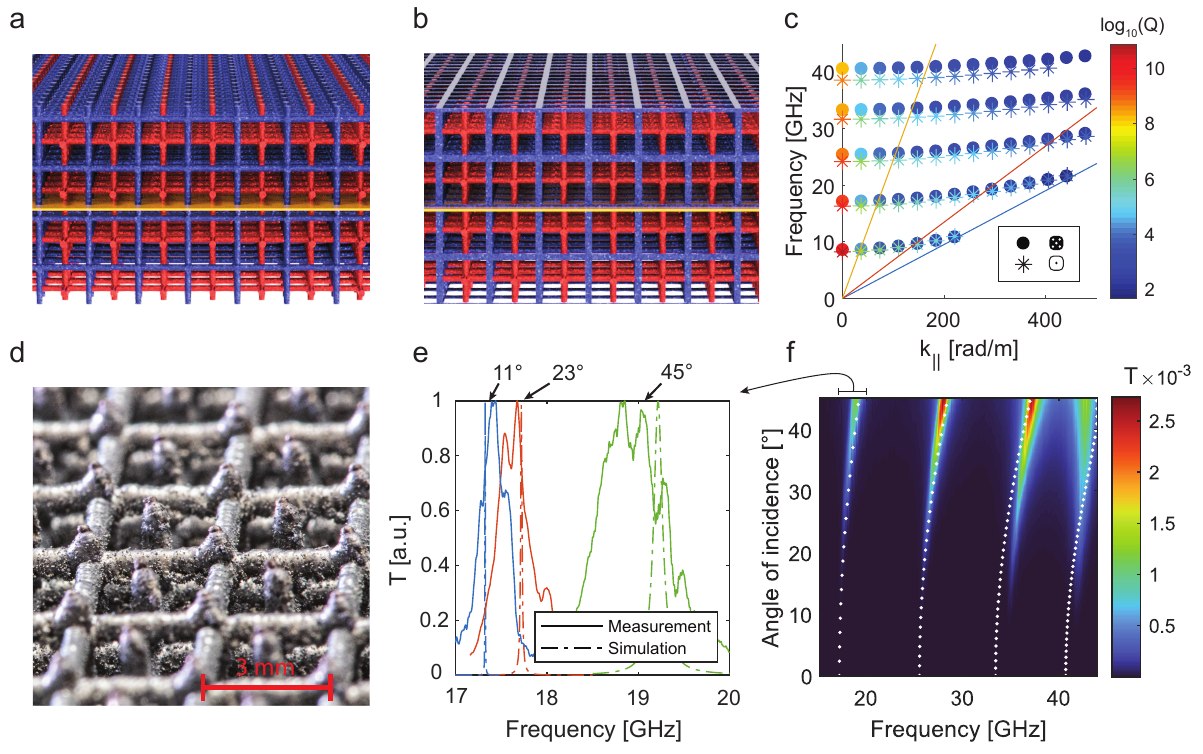}
\centering
\caption{\textsf{a} Model rendering of a PDN slab with protruding \emph{rods} at the interface ($\rods$). The slab is treated as infinite in the lateral dimensions by periodic boundary conditions. \textsf{b} Same as \textsf{a} but with a quarter unit cell shift in all dimensions resulting in a \emph{tennis net} morphology at the interface ($\tn$). \textsf{c} Slab quasi-normal mode band structure for the two geometries shown in \textsf{a} \& \textsf{b} with lattice constant $a\eq\SI{3}{\milli\metre}$. Each point represents a solution to the frequency domain eigenproblem at fixed Bloch wave number $k_\parallel$, and the colour encodes the logarithmic quality factor ($\log_{10}(Q)$). The solid lines show the free space dispersion for plane waves at \SI{90}{\degree} (blue), \SI{45}{\degree} (orange) and \SI{11}{\degree} (yellow) angles of incidence. The frequency of the fundamental mode at normal incidence is \SI{8.24}{\giga\hertz} for $\rods$ and \SI{8.67}{\giga\hertz} for $\tn$. \textsf{d} Photograph of the top surface of a metallic PDN slab with protruding \emph{rods} termination ($\rods$) used for the measurements. An image of the full 60\texttimes60\texttimes4 unit cell sample is shown in \figref{fig:supp:full_sample} in the supplementary information. \textsf{e} Individual second harmonic measurements compared to the simulation data for excitation at angles of incidence $\theta\eq\pi/4$, $\pi/8$, and $\pi/16$ (\SI{45}{\degree}, \SI{23}{\degree}, and \SI{11}{\degree}). \textsf{f} Heatmap of the measured transmission for the sample shown in \textsf{d}, superimposed with the expected peak position calculated in \textsf{c} (white dots).}
\label{fig:nets-and-resonances}
\end{figure}

Due to the difficulty fabricating PDNs, research on this topic has remained theoretical until recent years.
Since the advent of 3D printing, or more specifically additive manufacturing methods such as selective laser melting, it has become much easier to fabricate these metamaterials with arbitrary geometries at microwave length scales, facilitating experimental investigations.
Light tunnelling in PDNs was first experimentally demonstrated in 2021 \cite{Powell2021}.
The theoretically infinite finesse at normal incidence, where the vacuum field is decoupled from the EAW modes within the PDN (called dark modes in the paper), was significantly reduced by introducing an array of protruding L-shaped antennas on each side of the slab.
In stark contrast, a PDN with an unmodified tennis net termination surface (\figref{fig:nets-and-resonances}\textsf{b} shows only extremely weak light tunnelling at the quasi-BIC bands, even for large angles of incidence above \SI{45}{\degree} \cite{Dong2023}.

We show here that these recent experimental findings cannot be explained by previous homogenisation models, which are independent of the PDN termination plane.
Instead, the surface topology of the PDN termination profoundly influences the boundary conditions for the EAW bulk states, significantly changing the frequency and quality factor of the quasi-BIC bands.
These findings pave the way for a better understanding of PDN materials and their unique way of transporting electromagnetic energy through the material at frequencies below the plasma frequency.
This understanding is essential to exploit the large, anisotropic and robust quality factors of PDNs for applications such as lensing \cite{Wang2021}, coherent light generation \cite{kodigala2017a}, second harmonic generation, and biochemical sensing \cite{xu2023}.

\section*{Results and Discussion}

\subsection*{Plasmonic Double Nets}
Bulk PDNs behave like a warm, non-interacting double plasma in the low-frequency limit, giving rise to the longitudinally polarised EAW band \cite{Wang2023}.
For PDNs with cubic symmetry, the EAW dispersion is linear and isotropic, a behaviour well known from Maxwellian materials such as photonic crystals or metamaterials based on topologically disconnected meta-atoms.
While for the non-interacting double plasma, all EAW fields vanish, the weak interaction between networks in true PDNs leads to a longitudinal polarisation, i.e.\ the average electric field over the unit cell points in the direction of the wave vector.

These fundamental properties of EAWs show strong resilience to perturbations such as different network sizes, varying network offsets, and perturbatively breaking the cubic symmetry of the unit cell \cite{Wang2023}.
In addition, since the longitudinal EAW band always originates at zero frequency, the robust and controllable single-mode behaviour with topological protection spans a wide frequency range bounded by the effective plasma frequency $\omega_{\rm{p}}$.
While close to the perfectly conducting limit, applicable up to mid-infrared frequencies, the effective plasma wavelength $\lambda_{\rm{p}}$ depends mainly on the lattice constant of the unit cell $a$, with $a \inl\lesssim \lambda_{\rm{p}}$. 
PDNs are, therefore, excellent optical functional materials because their properties are extremely robust to fabrication imperfections and environmental fluctuations, for example, by introducing active components such as nonlinear materials or quantum emitters.

However, the experimental characterisation of PDNs is challenging.
A PDN slab supports BICs with infinite lifetime due to the orthogonality between the transverse electric field outside the PDN and the EAW inside.
External excitation of these BICs is naturally impossible.
Away from normal incidence, the electric fields outside and inside the material begin to overlap if the incoming wave is polarised in the plane of incidence (p-polarised).
Therefore, a small fraction of the incoming plane wave energy can couple into the network structure if it is spectrally close to the associated quasi-BIC resonances.
Previous homogenisation models and conventional wisdom in the metamaterials field suggest that the strength of this in-coupling depends mainly on the constituent materials and geometrical parameters of the bulk structure.
In contrast, recent studies on gyroid single-net metamaterials have revealed a strong polarisation dependence on the termination of the metamaterial \cite{Dolan2019,Cote2022}.
Here, we show that the quality factor of the quasi-BIC states in PDNs, and thus the coupling of light into the structure, changes by orders of magnitude depending on the metamaterial termination.

\subsection*{Impact of termination planes on resonant frequencies}

We focus here on the body-centered cubic pcu-c double-net \cite{okeeffe2008} (\figref{fig:nets-and-resonances}\textsf{a,b} that has been mostly studied in previous works both theoretically \cite{Shin2007a,Latioui2017,Wang2023} and experimentally \cite{Powell2021,Dong2023}.
It consists of two intertwined Cartesian lattices that are shifted by half a unit cell in all three directions with respect to each other.
We consider a $[001]$ inclination of the terminating slab surface, referred to in this paper as the $z$ direction.
Since the pcu-c double-net is a body-centred cubic structure, distinct terminations can be observed for the termination parameters $\tau\inl\in\{0,\text{\textonequarter}\}$ where we adopted the definition from \cite{Dolan2019}.\footnote{That is, $\tau$ is in units of the structural periodicity (here the lattice constant) and $\tau\inl=0$ parameterises a 24h Wyckhoff plane passing through a 2a Wyckhoff point at the crystallographic origin \cite{aroyo2016}.}
$\tau\eq0$ (symbolised as \tn{} here) gives rise to a \emph{tennis net} topology of one network as illustrated in \figref{fig:nets-and-resonances}\textsf{b} and $\tau\eq \text{\textonequarter}$ (symbolised as \rods{} here)  shows a protruding \emph{rods} pattern in both networks, as seen in \figref{fig:nets-and-resonances}\textsf{a}.
For two \replace{four unit cell thick}{four-unit-cell-thick} slabs with these reference terminations, we compute the slab mode band structure by full-wave simulations.
These are the quasi-normal slab modes \cite{Lalanne2018} which satisfy the Sommerfeld radiation condition for $|z|\inl\to\infty$ and are subject to Floquet periodic boundary conditions with variable in-plane wave number $k_x\eq\frac{\omega}{c}\sin(\theta)$ and $k_y\eq0$, where $c$ is the speed of light in vacuum and $\theta$ is the polar radiation direction in the far-field.

\figref{fig:nets-and-resonances}\textsf{c} shows the slab dispersion $f(k_x)\eq\frac{\omega(k_x)}{2\pi}$ for both terminations for a microwave PDN with lattice constant $a\eq\SI{3}{\milli\metre}$ and strut radius $r\eq\SI{0.2}{\milli\metre}$.
The two terminations evidently lead to different resonant frequencies of the slab modes even at $\theta\eq0$.
This contradicts the established picture where the PDN homogenises to effective parameters.
In the normal radiation direction, the PDN is decoupled from the outside and behaves like an optical cavity with an infinite quality factor.
In the homogenised picture, this corresponds to hard wall boundary conditions, for which the $z$-currents vanish at the interface.
This leads to BICs with a simple resonance condition \cite{Wang2023}
\begin{equation}\label{eq:FSR}
    f_{\rm{BIC}} = n\Delta f\quad\text{, with}\quad n\in\mathbb{N} \quad\text{and}\quad \Delta f = \frac{\kappa c}{2l}\,,
\end{equation}
where $l$ ($4a$ in our case) is the slab thickness and $\kappa\inl\approx0.7$ is the plasma pressure parameter weakly dependent on the strut diameter so that $\kappa c$ is the EAW's wave velocity.
For the lowest BIC resonances, the theory agrees reasonably well with the resonances of the protruding rod termination but deviates from the resonance of the tennis net termination.
The theory generally overestimates the higher-order resonances because the effective medium model does not consider the finite size of the Brillouin zone and, therefore, overestimates the group velocity of the EAW at frequencies closer to the effective PDN plasma frequency.
The significant termination dependence of the resonance frequency contradicts conventional wisdom and indicates that PDNs are not yet well understood under realistic experimental conditions.
In the following, we show that the termination also strongly influences the lifetime of the radiating quasi-BIC modes at finite angle $\theta$.

\subsection*{Extreme termination dependence of quasi-BIC lifetimes}

The quality factor $Q\deq\Re\{f\}/\Im\{f\}$ quantifies the lifetime of the quasi-normal slab modes.
\replace{While}{Under theoretical conditions (ideal geometry, plane wave excitation, PEC network),} it diverges for the BIC states at normal radiation direction for all terminations\replace{,}{. At finite radiation angles,} different terminations have significantly different $Q$ factors\delete{ at finite radiation angles}.
Indeed, \replace{the imaginary part of the quasi-BIC frequency $\Im\{f\}$,}{the} colour-coded \add{$Q$ factor shown} on a logarithmic scale in \figref{fig:nets-and-resonances}\textsf{c}, depends strongly on the choice of termination and differs by two orders of magnitude for the fundamental mode at an incidence angle of $\theta\eq \pi/16$ (see Tab.~\ref{tab:Q_F} for details).
The rod termination $\rods$ generally gives rise to a lower $Q$ factor than the tennis net termination $\tn$.
While a high $Q$ factor is desirable for most applications, optical characterisation by transmission experiments benefits from a lower $Q$ factor because the quasi-BIC transmission linewidth through the PDN slab is inversely proportional to the $Q$ factor.
Experimental characterisation of these \add{extremely} high-$Q$ resonances by excitation and measurement in the far field outside the PDN is virtually impossible in a slab
\replace{because the geometry does not provide a stable resonant cavity \cite{Saleh2019}. 
Instead, the two planar interfaces of the slab create a metastable resonator configuration, leading to divergent behaviour even with the slightest imperfections within the measurement setup.
This fundamental problem is compounded by manufacturing imperfections, such as defects or random curvature at the termination plane, and practical limitations, such as a finite slab area.
}{
due to fabrication imperfections, setup limitations and plasmonic losses.
}

\begin{table}[t]
    \begin{center}
    \begin{tabular}{|l|c|c||c|c|}
    \hline
            & Termination   & $\theta$      & $\Re(f)$ [\si{\giga\hertz}]   & $\log_{10}(Q)$ \\\hline
        \multirow{2}{8em}{Homogeneous Plasma Model}  & \multirow{2}{2em}{NA} & 0         & 8.75  & NA    \\
                                &                       & $\pi/16$  & 8.92  & NA    \\\hline
        \multirow{4}{8em}{Scattering Theory}    & \multirow{2}{2em}{\tn} & 0    & 8.19  & NA  \\
                                &                   & $\pi/16$  & 8.34  & 5.1 (7.5)   \\
                                & \multirow{2}{2em}{\rods}  & 0 & 8.57  & NA          \\
                                &                   & $\pi/16$  & 8.74  & 4.4 (4.1)   \\\hline
        \multirow{4}{5em}{Simulations}    & \multirow{2}{2em}{\tn} & 0    & 8.23  & NA  \\
                                &                   & $\pi/16$  & 8.30  & 6.5   \\
                                & \multirow{2}{2em}{\rods}  & 0 & 8.65  & NA      \\
                                &                   & $\pi/16$  & 8.70  & 4.3   \\\hline
        Microwave Experiments   & \multirow{1}{2em}{\rods} & $\pi/16$  & 17.4 (8.7)  & 2.0    \\\hline
    \end{tabular}
    \end{center}
    \vspace{1em}
    \caption{Comparison of resonant frequency ($f$) and quality factor ($Q$) for the resonant slab mode in PDNs from different approaches: The hydrodynamic plasma model \cite{Wang2023}, our new model, full-wave simulations, and microwave experiments. We distinguish between the tennis net $\tn$ and the rods termination $\rods$ and consider normal radiation $\theta\eq0$ and radiation at a small angle $\theta\eq\pi/16$. Two quality factors have been extracted from the semi-analytical model based on the frequency definition $Q\eq\frac{\Re f}{\Im f}$ and the field energies (in brackets), see Section \ref{sec:supp:quasiBIC}. The measurement shows the value of the first harmonic instead of the fundamental mode due to setup limitations, with the extrapolated fundamental mode frequency in brackets.}
    \label{tab:Q_F}
\end{table}

The first experimental work on the subject \cite{Powell2021} \replace{introduced}{circumvented these problems by introducing} antennas on the termination surface of a $15\times15$ unit-cell slab to couple light in and out of the metamaterial, artificially lowering the $Q$ factor of the BIC states to experimentally convenient values.
This enabled the measurement of the otherwise \emph{dark} modes within the slab.
In stark contrast, the first experimental observations of an unmodified PDN slab \cite{Dong2023} showed only a very weak transmission signal only at large angles of incidence above ${\approx}\SI{45}{\degree}$, despite the fabrication of a large high-quality sample ($50\times50$ unit cells) and the use of a sophisticated setup.
The absence of transmission lines at low angles of incidence is due to the sample having a tennis net termination plane, which leads to very high $Q$ factors even at large angles of incidence, as predicted by our simulations, see \figref{fig:nets-and-resonances}\textsf{c} \& \figref{fig:Q_eval}\textsf{a}.

On the contrary, we \add{theoretically} predict more than two orders of magnitude lower quality factors for a rod-terminated PDN slab.
To show that this extreme termination dependence is also observed under experimental conditions, we fabricated a $60\times60\times4$ unit cell PDN slab using a selective laser melting 3D printer.
The cubic lattice constant of the pcu-c PDN was set to $a\inl=\SI{3}{\milli\metre}$ and the strut radius to $r\inl=\SI{0.2}{\milli\meter}$.
Using a standard two microwave horn setup in combination with a vector network analyser (see \figref{fig:supp:microwave}, further details in the Methods section), we measured the transmissivity between 0 and \SI{45}{\degree} angle of incidence in a frequency range between 16 and \SI{44}{\giga\hertz}, as shown in \figref{fig:nets-and-resonances}\textsf{f}.
The position of the experimental transmission lines is in very good agreement with the simulated quasi-BIC bands (dots).
\add{
The $Q$ factor and the signal quality are, on the other hand, still greatly reduced. In particular, the experiments suffer from
\begin{enumerate}
    \item Fabrication imperfections: The sample shows a small global bending due to the thermal stress during the fabrication, slightly visible in \figref{fig:supp:full_sample}. This implies that the termination of the metamaterial is slightly curved. The network surface also exhibits a significant surface roughness, as clearly visible in \figref{fig:nets-and-resonances}\textsf{d}. While this roughness is much smaller than the resonance wavelength, it still influences the current flow in the wires and changes their electric properties. Finally, there are a number of surface defects at the top of the metamaterial in the form of missing struts, see \figref{fig:supp:defects}. These technically perturb both the translation and the point symmetries of the slab, on which the prediction of the BICs is based.
    \item Setup limitations: While the sample has been surrounded by absorbing foam, it is still finite in the $x-y$ plane (60x60 unit cells), allowing scattering loss of light away from the receiving horn. More importantly, the illumination source is a highly directional microwave horn, but not a perfect plane wave. In the far field, it produces a slightly diverging beam that closely resembles a zero-order Hermite Gaussian beam, with an opening angle of around \SI{5}{\degree}, causing a lineshape broadening, largely independent of the incidence angle.
    \item Plasmonic losses: While metals at microwave frequencies are generally very good conductors, the PEC approximation is not a good description when dealing with ultrahigh $Q$ factors, particularly for the aluminium alloy that our sample is made of. We have simulated the bulk EAW wave at the resonance frequency, replacing the PEC boundary conditions with impedance boundaries for the complex Drude permittivity of aluminium, and obtained a non-radiative $Q$ factor below $1000$. This suggests that the dissipative loss channel is, indeed, the limiting factor. 
\end{enumerate}
While the robustness of the BIC resonances with regard to fabrication imperfections has been studied in \cite{Wang2023}, we will address the influence of the microwave horn antennas and the plasmonic losses in more detail in the discussion below.
}
\delete{
This is even though the 3D printed metal surface has a rough surface as seen in \figref{fig:nets-and-resonances}\textsf{d}, and the sample has defects such as a slight curvature of the surface due to thermal stress during printing, as well as a small number of missing struts on the back of the sample (see \figref{fig:supp:defects}).}
\delete{This demonstrates provides evidence for the resilience of the EAW-induced quasi-BIC bands to geometrical perturbations as predicted in \cite{Wang2023}}
Due to the significantly reduced \add{theoretical} $Q$ factors compared to the sample in \cite{Dong2023}, we \add{still} detected the resonant mode transmission lines
\replace{with angles of incidence of down to 10 degrees.}{down to very small angles of incidence}
\delete{However, the experimental linewidth is much larger than predicted by the simulation}
, as shown in \figref{fig:nets-and-resonances}\textsf{e} for $\theta\eq\SI{10}{\degree}$, \SI{22}{\degree}, and \SI{45}{\degree}. 
\delete{This corresponds to an experimental reduction in the $Q$ factor due to the aforementioned fabrication constraints (finite slab size and an unstable cavity) and perturbations (curved termination surface, metal roughness, defects), as well as the microwave measurement itself.
Although the sample was embedded in an absorbing foam, as in \cite{Dong2023}, to avoid capturing diffraction around the sample, the horns had to be placed optically close to the sample (emitting horn much closer than in \cite{Powell2021}) to increase the signal-to-noise ratio (details in the Methods section).
This implies that the excitation was only a rough approximation of the theoretically assumed plane wave, while the receiving horn integrated over a non-negligible solid angle.}

\subsection*{Predicting the resonance frequency shift}

\begin{figure}[tbp]
\includegraphics[width=\textwidth]{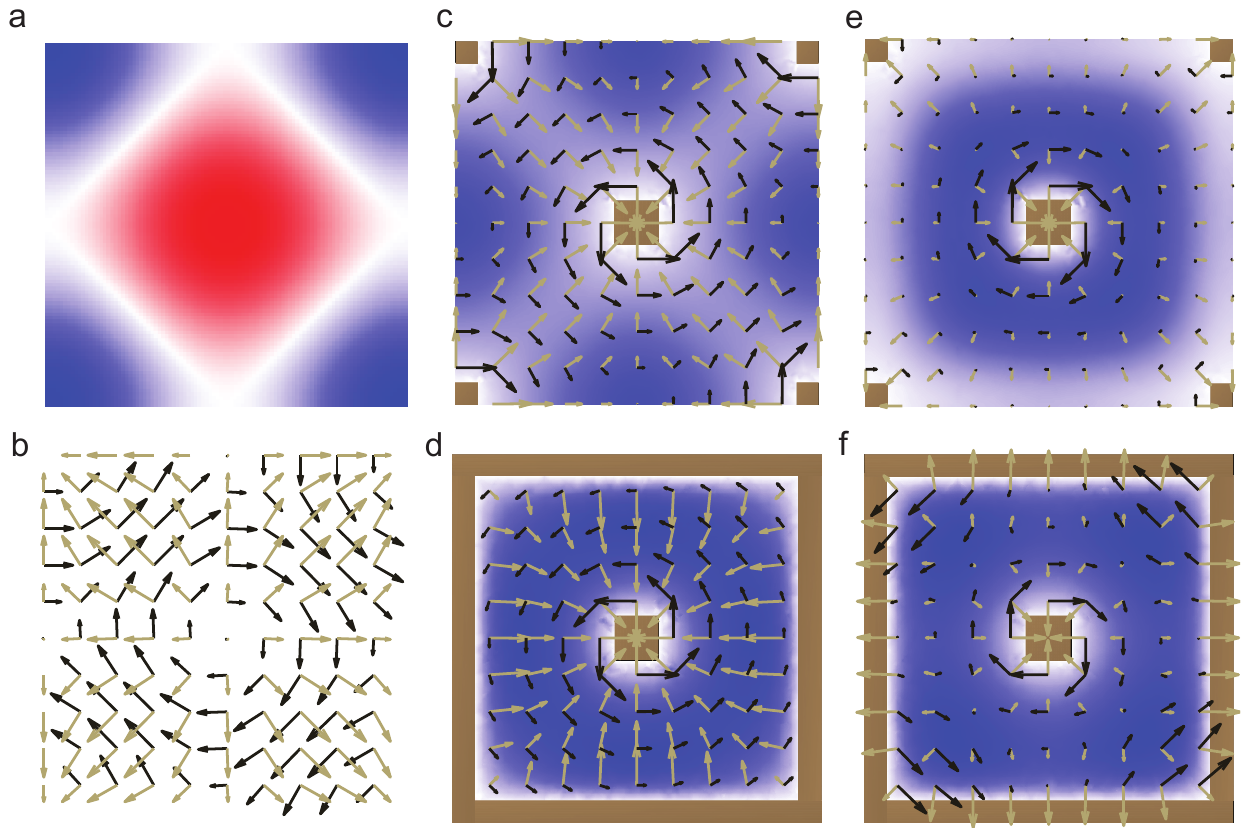}
\centering
\caption{Analytical and simulated field patterns used to predict the frequency and quality factor shift between different PDN terminations. The diverging heat maps represent the normalised out-of-plane electric field intensity $E_z$ (red positive and blue negative field). The light brown and black vectors represent the in-plane electric and magnetic fields, respectively. For all fields, we have chosen a phase in the optical cycle where these are large. \textsf{a},\textsf{b} First Bragg order basis fields in the group theoretical channel of the EAW, \emph{cf.}~\ref{tab:surf_basis} in the supplementary materials: \textsf{a} Out-of-plane electric field, \textsf{b} in-plane E and H fields in the first Bragg order. \textsf{c}-\textsf{d} The electromagnetic fields for an EAW extracted from a \emph{bulk} simulation at the \textsf{c} \rods{} and \textsf{d} \tn{} termination. \textsf{e}-\textsf{f} Electromagnetic fields for an evanescent longitudinal mode with imaginary wave number $k_z \eq 1.03 [\mathnormal{i}\frac{2\pi}{a}]$ extracted from a \emph{bulk} simulation at the \textsf{e} \rods{} and \textsf{f} \tn{} termination.}
\label{fig:fields}
\end{figure}

Here, we introduce a novel approach to predict the frequency shift between the two terminations observed in the simulations.
This shift can be predicted by coupling the bulk PDN modes with the vacuum Rayleigh basis \cite{rayleigh1907}.
The novelty of this method lies in the consideration of higher Bragg orders (both in propagation and in the lateral direction).
\replace{In particular, we employ an interface field-matching approach that expands the fields at the PDN-vacuum interface into the bulk subspace modes, including higher-order evanescent states.
We then match the field components within the group-theoretically allowed scattering channel at the interface by linear superposition of the electromagnetic bulk modes present on both sides of the boundary. In particular, the }
{
In particular, we employ a field-matching approach that goes beyond the homogenized model \cite{Latioui2017}. We first expand the fields in the different domains (vacuum and slab) into a set of propagating and evanescent Floquet states. This idea was first used to solve the scattering problem of 2D photonic crystals around 2010 \cite{lawrence2009,lawrence2010}. A commercial software based method to compute the associated evanescent states for 3D structures was introduced shortly after \cite{fietz2011}, highlighting the pivotal role of these states in expanding both localised and delocalised states at photonic crystal and metamaterial interfaces. In 2015, we developed a plane-wave-based method to calculate evanescent Floquet modes of 3D photonic crystals, and showed how these states can be used to expand the fields inside a slab geometry \cite{cryst5010014}. The convergence behaviour of this Fourier-based expansion is only linear due to Gibb's ringing.
We here, however, only use a first-order expansion of the fields to obtain a correction to the termination-independent homogenised model to obtain physical insight into the termination dependence. In contrast to the previous literature, we exploit the rotational symmetry of the problem to further reduce the dimensionality of the scattering problem. The
}
slab exhibits a $\mathbf{C_{4v}}$ point symmetry that allows the observed field patterns in a cross-section of the bulk unit cell to be mapped to the trivial irreducible representation of the EAW within this group, see supplementary material for theoretical details.
The electromagnetic field components inside the PDN are projected from a \emph{bulk} simulation with Floquet boundary conditions at the two different terminations \tn{} and \rods.
The frequency $f\,{\approx}\,\SI{8.1}{\giga\hertz}$ was chosen to match the fundamental slab resonance for the normal radiation direction with the network parameters described in the text.
The vacuum projections, on the other hand, are naturally $z$-independent and have an analytical expression.
The in-plane field patterns used for this projection are shown in \figref{fig:fields}\textsf{b}.
The electric field exhibits a radial and the magnetic field a curling pattern, as derived in Section \ref{sec:supp:symmsurf} in the supplementary material.
At vanishing $k_\parallel$, the simulations indicate a diverging $Q$ factor, which stems from the absence of longitudinal plane waves in the scattering channel that can radiate the energy away from the slab.
By definition, the longitudinal slab modes do not have a homogeneous in-plane field pattern, so we need to match the radial E-field, the curled H-field, and the homogeneous out-of-plane E-field.
To form a well-defined scattering problem in which the degrees of freedom match the interface conditions, we must consider three modes per interface: The EAW, a longitudinal evanescent mode in the slab, and a first-order Bragg mode outside the slab, see Section \ref{sec:supp:expansion} for details.
Let us now define the 3\texttimes3 matrix $\mat{M}$, where each column describes a mode and each row the projected field components of these modes. 

For the EAW, we have to consider both propagating and counter-propagating waves that create a standing wave in the slab.
Both are related by the symmetry of the slab, which is mirror symmetric for the tennis net termination and features a glide-mirror symmetry in case of the rods termination, see \figref{fig:nets-and-resonances} \textsf{a,b}.
This introduces a different phase for the different field components and terminations; see supplementary Section \ref{sec:supp:BICfreqs}.
The frequency is found by tuning the accumulated phase of the counter-propagating wave $p=\exp(i\cdot k_z\cdot N\cdot a)$, where $N$ is the number of unit cells in $z$, leading to the following generalised eigenproblem
\begin{equation}
    \mat{M}\cdot\mathbf{v} = p\cdot\begin{bmatrix}
        C_{1} & 0 & 0  \\
        C_{2} & 0 & 0 \\
        C_{3} & 0 & 0 
    \end{bmatrix}\cdot\mathbf{v},
    \label{eq:eigenproblem}
\end{equation}
where $C_i$ are the symmetry copies of the EAW field projections and the eigenvector $\mathbf{v}$ stores the bulk mode (EAW, evanescent, and first Bragg order) coefficients, while the phase $p$ takes the role of an eigenvalue.
The frequency is finally obtained using the EAW bulk dispersion $\omega \eq \kappa{\cdot} c {\cdot} k_z$, where the plasma pressure parameter $\kappa\,{\approx}\,0.7$ is obtained from the bulk bandstructure dispersion of the EAW.

\subsection*{Predicting the resonance lifetime}
At non-normal radiation, the trivial $C_{4v}$ scattering channel weakly couples to the transverse scattering channel of the radiating $p$-polarised vacuum plane wave, with finite $E_x$ and $H_y$ components in the zero Bragg order.
The eigenproblem \eqref{eq:eigenproblem} becomes a 5\texttimes5 problem and contains the radiating plane wave and a transverse evanescent wave inside the slab in addition to the three original modes.
Due to the weak coupling, the problem separates into the original 3\texttimes3 block and the transverse 2\texttimes2 block, which are coupled through off-diagonal blocks that are obtained by perturbation theory in linear order for small $\rm{d}\theta$ from the original fields.
This corrects the phase $p\approx p_0+p_1\rm{d}\theta^2$, where $p_0$ is the normal radiation phase, and we derive a closed form for $p_1$ in Section \ref{sec:supp:quasiBIC}.
The corresponding change in the real part of the frequency is negligible compared to the correction from the EAW dispersion $\omega\eq \kappa{\cdot} c {\cdot} k_z /\cos(\rm{d}\theta)$.
However, the correction in the imaginary part of the frequency in second order in $\rm{d}\theta$ gives rise to a finite quality factor $Q\eq Q_1{\cdot}\rm{d}\theta^{-2}$, where we derive a frequency-based and a field-based expression for $Q_1$ in Section \ref{sec:supp:quasiBIC}.
The resulting quality factors agree well with the simulation results for the \rods{} termination; see Table \ref{tab:Q_F}.
For the \tn{} termination, the frequency-based definition underestimates the $Q$ factor, while the field-based definition overestimates it.
In the full theory, both should be identical due to energy conservation, but the semi-analytical model assumptions are less well justified for \tn.
For example, the surface field basis does not consider the exclusion volume of the struts, and we assumed the fields to vary according to the Floquet dispersion in the $z$ direction.
\begin{figure}[t]
    \centering
    \includegraphics[width=\linewidth]{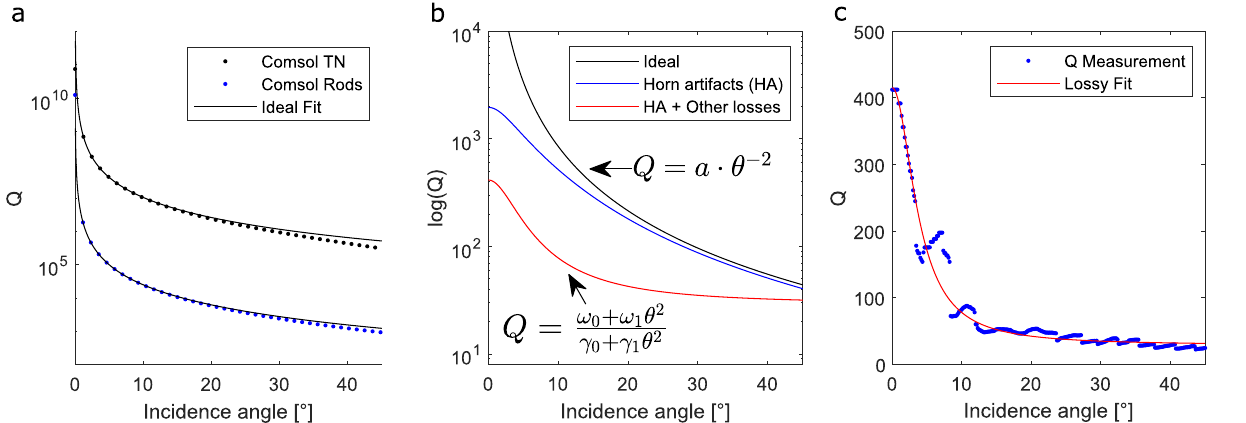}
    \caption{\add{Comparison of the extracted quality factors ($Q$) from the simulated and measured data. \textsf{a} Quality factor as a function of incidence angle obtained from simulations using the relation $Q\eq\Re f/\Im f$. The data shown corresponds to the fundamental mode around \SI{8}{\giga\hertz}. It was fitted with a curve of the form $Q\eq Q_0\cdot \theta^{-2}$, where the fit of the tennis-net terminated slab exhibits $Q_{0\mathrm{TN}}\eq \si{1.05e9}$ (in units of squared degrees), and the rods terminated slab $Q_{0\mathrm{Rods}}\eq \si{2.51e6}$. \textsf{b} Schematic $Q$ relationship in the ideal system for the first excited resonance (black), together with the horn source corrected shape (red), and the lossy system used to fit our data (blue). \textsf{c} Extracted $Q$ values from the measurement data, obtained by using the FWHM of the peak in each successive measurement plot and using the relation $Q = f/\Delta f$. The data corresponds to the lowest mode measured around \SI{16}{\giga\hertz}. It was fitted with the expected curve form for lossy systems shown in \textsf{b}.}}
    \label{fig:Q_eval}
\end{figure}

\add{
A simple perturbation theoretical argument -- without the need for a complex scattering theory -- explains and connects the simulated (ideal case) and experimental results with a greatly reduced signal strength and $Q$ factor.
In the ideal case, the $\theta^{-2}$ scaling law is a consequence of the BIC frequency perturbation when going away from normal radiation $\omega(\theta) \eq \omega_0 {+} (\omega_1+i\gamma_1)\theta^2$.
Note that the first-order perturbation in $\theta$ vanishes due to reciprocity, even if the mirror symmetry is broken.
A similar behaviour is caused by any perturbation that is analytical and symmetric at the evaluation point.
Examples in the literature include geometric perturbations \cite{koshelev2018} and plasmonic non-localities \cite{liang2024}.
Thus, the quality factor is described by $Q\inl\approx \frac{\omega_0}{\gamma_1}\theta^{-2}$, with a single free fit parameter $\alpha\deq \frac{\omega_0}{\gamma_1}$.
As seen in \figref{fig:Q_eval}\textsf{a}, the simulated curves follow this functional relation extremely well, only deviating significantly for large angles.
}

\add{
Ideal conditions, however, imply plane wave excitation, while excitation with real sources significantly reduces the quality factor.
Particularly for focused light, this is a limiting factor that has recently been mitigated by combining two different BICs in a metasurface \cite{liang2021}.
The theoretical quality factors for PDNs stay very high (above 100) even for large angles of incidence, so that we do not require such an elaborate design.
Nevertheless, the microwave horn antenna provides a far field profile that resembles a Gaussian beam with a small, but finite, opening angle around \SI{5}{\degree}.
It, therefore, probes quasi-BIC states over an angular range, resulting in a convolution of the angle-dependent spectrum with a normal distribution.
This leads to a quadratically additive lineshape broadening $\sigma^2\eq\gamma_1(3\theta_{\rm{h}}^4{+}6\theta_{\rm{h}}^2\theta^2{+}\theta^4)$ and a $Q$-factor of
\begin{align*}
    Q = \frac{\omega_0}{\gamma_1\sqrt{3\theta_{\rm{h}}^4{+}6\theta_{\rm{h}}^2\theta^2{+}\theta^4}}
\end{align*}
shown in red in \figref{fig:Q_eval}\textsf{b}.
The quality factor limitation due to the source at normal incidence solely depends on the ideal fit parameter $\alpha$ and the source's numerical aperture.
The angle-widening generally induces a frequency perturbation that is non-analytical in $\theta_{\rm{h}}$ and $\theta$ at $\omega_0$ because of the root function.
In contrast, fabrication imperfections and plasmonic losses are expected to affect the eigenfrequency analytically.
This leads to the general expression
\begin{align*}
    Q = \frac{\omega_0 + \omega_1 \theta^2}{\gamma_0 + \gamma_1 \theta^2}
\end{align*}
with three independent fit parameters.
This expression describes the measured data in \figref{fig:Q_eval}\textsf{c} well, and is reproduced in \figref{fig:Q_eval}\textsf{b} for comparison.
The wavy distortions in the experimental data stem from a Fabry-P\'erot interference pattern seen in \figref{fig:nets-and-resonances}\textsf{e} that is likely caused by the two microwave horns forming a large cavity around the metamaterial.
This pattern is multiplied by the expected Lorentzian line shape and influences the estimated linewidth periodically.
The theoretically estimated dissipative $Q$ for the experimental aluminium alloy double net is on the same order of magnitude as the measured $Q$ in \figref{fig:Q_eval}\textsf{c}.
This strongly indicates that dissipation is indeed the dominant factor that limits the resonance lifetime, while fabrication and setup imperfections play a minor role.
While this is an obvious fact for BICs in plasmonic metamaterials at higher frequencies \cite{liang2024}, it demonstrates that the general assumption to approximate metals through PECs at microwave frequencies \cite{collin2001,smith2000,garciadeabajo2005} is flawed for high-$Q$ BIC applications.
}

\replace{
Nevertheless, the model qualitatively predicts the much higher $Q$ factors for \tn{} terminated slabs well.
It traces them back to a negligible homogeneous $E_z$ field of the EAW wave within the tennis net region and, consequently, a much weaker out-coupling into the propagating $p$-polarised radiating plane wave.
}{
The lifetime limitation by plasmonic dissipation explains why the signal-to-noise ratio deteriorates strongly for very small angles below \SI{10}{\degree} in our experiments, and why the experimental paper on \tn{} terminated slabs \cite{Dong2023} only produced suitable results for very large angles of incidence.
At the same time, the theoretical scattering model explains the origin of the radiative decay, and the strong discrepancy between the \tn{} and the \rods{} termination.
Radiation at a finite angle takes place in the zero Bragg order via the transverse $p$-polarised plane wave.
The perturbation theory requires a finite zero Bragg order, that is, a homogenised electric field $E_z$ at the termination plane at normal radiation, which couples into the lateral electric field of the $p$-polarised wave.
The $Q$ factor is, therefore, dominated by the ability of the PDN to generate a homogeneous field at the termination plane.
At low frequencies, the dominant EAW wave generates such a field only between two tennis nets at the rods termination, but not at the tennis net termination.
This can be seen from the quasistatic approximation \cite{chen2018}, in which the two networks are on opposite constant potentials.
For thin wires, two neighbouring tennis nets act as a capacitor with an associated electric field in between, but a vanishing $E_z$ in the plane.
Alternatively, one can see the tennis net as an array of very short square waveguides with a conducting centre strut.
These waveguides support a TEM mode with vanishing $E_z$ and zero cut-off frequency, but the first excited TM mode with finite $E_z$ has a cut-off wavenumber that is close to $2\pi/a$, ten times larger than the BIC resonance wavenumber.
All higher-order modes with finite $E_z$ are therefore strongly evanescent and impedance-mismatched with their environment.
The $z$ component of the electric field is thus strongly suppressed at the center of the tennis net domain.
}

\section*{Conclusion and Outlook}

The experimental and simulation data presented above clearly highlight the importance of the termination plane when considering resonant slab modes in PDNs.
The established effective medium models are insufficient to predict the behaviour of EAWs in a slab configuration.
We introduce a semi-analytical scattering theory that explains the strong influence of the surface termination and its geometry on the resonant frequencies and the associated quality factors.
The varying field contributions in the different Bragg orders of the surface unit cell for different terminations can substantially alter the impedance of the perfect resonator walls at normal radiation conditions.
In particular, the homogeneous $E_z$ component used for hard-wall boundary conditions in the established literature is very small for a tennis net termination, leading to a substantial correction of the fundamental resonance frequency from \SI{8.75}{\giga\hertz} to \SI{8.23}{\giga\hertz}.
While the evanescent bulk states in the PDN must be considered in the scattering problem and influence the scattering problem, the tiny $E_z$ component for the tennis net termination also qualitatively explains that the quality factors are more than two orders of magnitude higher for this termination compared to a rod terminated sample.
This is due to the cross-coupling in the zero Bragg order between the $E_z$ component of the longitudinal EAW wave and the $E_x$ and $H_y$ components of the radiating vacuum plane wave at small radiation angles in the $x{-}z$ plane.

Another qualitative explanation is based on the observation that EAWs are carried by currents along the $z$ wires (proportional to the curling H-field in \figref{fig:fields}).
In the case of the rod termination, these protruding struts can act like a dipole antenna array, fed by these currents, radiating energy away from the slab for finite wave vectors parallel to the slab.
For the tennis net termination, the dipoles sit below the percolating metallic structure in the termination plane that shields them from the outside.
This image is less rigorous but helps to intuitively understand the lower quality factor of the tennis net termination.
A similar explanation can be intuited from the same picture to explain the difference in resonant frequency between the two terminations.
We will first look at the rod termination, where the protruding struts act as a hard-wall boundary condition for the surface currents along the propagation direction.
The currents must decrease smoothly as they approach the interface so as not to accumulate a diverging charge density at the end of the small wires.
For the tennis net termination, the interconnected topology at the interface serves as a capacitance reservoir that allows the perpendicular currents to be redirected into the parallel struts.
This increases the apparent size of the cavity, resulting in a lower resonant frequency. 

\replace{Our work thus highlights the importance of the termination plane in studying resonant slab modes PDN materials and network-like metamaterials in general.}
{
Not only does this work highlight the importance of the termination plane in studying resonant slab modes in PDN materials and network-like metamaterials in general.
It also reveals the pivotal role of plasmonic losses in high-$Q$ microwave metamaterials.
While the PEC approximation is useful to generate a general physical understanding, it is not sufficient to obtain a reasonable estimate for the resonance position and linewidth, which is here limited by dissipative broadening.
}
Consciously choosing the termination \add{as demonstrated through microwave experiments with a 3D-printed PDN and using a good conductor such as silver as plasmonic material greatly} facilitates the experimental study of quasi-BIC bands down to small angles of incidence and improves the cavity performance.\delete{as demonstrated through microwave experiments with a 3D-printed PDN.}
Building on previous work by different groups, this research paves the way for a better understanding of PDN structures and their unique way of propagating energy below their effective plasma frequency through an electron acoustic wave.
Although this work introduces new tools for understanding and quantifying finite PDN structures, further research is needed to fully understand slabs and 3D resonators with varying crystal inclinations.

\section*{Materials and Methods}
Our slab PDN sample was printed using a SLM250HL selective laser melting machine and an AlSi10Mg powder from Carpenter Additive at the Swiss Innovation Park Biel/Bienne (SIPBB). It contains $60\times60\times4$ unit cells with a unit cell size of \SI{3}{\milli\metre} ($180\times180\times12$\,mm).

The experiments used a Keysight P5007A two-port vector network analyser (VNA) and LB-180400-25-C-2.4F antennas from A-Info Inc. A picture of the setup can be found in \figref{fig:supp:microwave}. The experimental data were obtained by placing the PDN sample on a rotating remote-controlled platform. The transmitting horn antenna was fixed at a distance of \SI{500}{\milli\metre}. The receiving horn antenna, also fixed, was kept close to the sample to reduce measurement noise from diffraction and reflections from the table. This was further improved by embedding the sample in microwave absorbing foam. Sample rotation and VNA sweep were remotely controlled by a MATLAB script. To ensure proper alignment of the setup, 601 discrete angles between --45\textdegree \space and 45\textdegree \space were measured. The 0\textdegree \space position was chosen as the symmetry axis in the acquired data. Each discrete angle measurement contained 20,001 sampled frequency data points between 4.4 -- \SI{44}{\giga\hertz} and was averaged over 200 sweeps to reduce noise. Data below \SI{16}{\giga\hertz} is not shown as it is outside the radiation range of our antennas.  
Simulations were performed with the COMSOL Multiphysics simulation software.

\section*{Acknowledgements}
We would like to thank Viola V.~Vogler-Neuling for helping with the goniometer setup. This work was supported by the Swiss National Science Foundation through Grant 200020\_188647 and by the Adolphe Merkle Foundation.

\printbibliography

\end{document}


\maketitle

\renewcommand{\thesection}{S1}
\section{Theoretical Details}

To understand the termination-dependent scattering physics at the slab, a fully homogenised approach \cite{Latioui2017} is naturally insufficient. Instead, we expand the fields at the interface within the irreducible point group representations of the slab geometry to efficiently describe and understand the termination-dependent scattering.

\subsection{Symmetry classification of the fields}\label{sec:supp:symmsurf}
The BIC states at normal irradiation ($\rm{d}\theta\eq0$) are mainly carried by two counter-propagating \emph{longitudinal} EAWs. For the $\{100\}$-oriented slab under consideration, we can classify the BIC state with respect to the underlying $C_{4v}$ point group with the character table shown in Table \ref{tab:C4v}.
The electric field of the EAWs transforms trivially (irreducible representation or \emph{irrep} $A$), while the magnetic field transforms trivially under proper rotations and acquires a character $\chi\eq{-}1$ under mirrors ($B_1$ irrep).

We consider a small lateral dispersion for the modal solutions within the different domains ($\rm{d}\theta\inl\ll1$). The fields are assumed to be periodic without the Bloch phase, so Floquet-periodic boundary conditions are intrinsically enforced (the electric field, for example, is $\vec{E}\exp\{\imath\,\vec{k}\cdot\vec{r}\}$). The modes in the slab and the semi-infinite vacuum domain outside are thus described by Maxwell's equations (in Gauss units),
\begin{equation}
    \underbrace{\begin{bmatrix}
        -\varepsilon\identity & \frac{\imath}{k_0}\,\mat{C} \\
        -\frac{\imath}{k_0}\,\mat{C} & -\identity
    \end{bmatrix}}_{\mat{M}_0}
    \underbrace{\begin{pmatrix}
        \vec{E} \\ \vec{H}
    \end{pmatrix}}_{\vec{F}} = 
    d\theta \underbrace{\begin{bmatrix}
        0 & \mat{X} \\
        -\mat{X} & 0
    \end{bmatrix}}_{\mat{M}_1}
    \vec{F}
    \text{.}
    \label{eq:MWE_modal}
\end{equation}
Here, the vacuum wavenumber is $k_0\eq\omega/c$, the small radiation angle is $d\theta\inl{\approx} k_x/k_0\inl{\ll}1$, and the permittivity $\varepsilon$ is identical $1$ in free space and is understood in the limit to ${-}\infty$ in the PEC struts. The curl $\mat{C}$ and the operator $\mat{X}$ are expressed by
\begin{equation}
    \mat{C} := \begin{bmatrix}
        0 & {-}\partial_z{-}\imath k_z & \partial_y \\ \partial_z{+}\imath k_z & 0 & {-}\partial_x \\
        {-}\partial_y & \partial_x & 0
    \end{bmatrix} \quad\text{and}\quad
    \mat{X} :=
    \begin{bmatrix}
        0 & 0 & 0 \\ 0 & 0 & -1 \\ 0 & 1 & 0
    \end{bmatrix}\,.
    \label{eq:curl}
\end{equation}

In \eqref{eq:MWE_modal}, the curl transforms similar to the magnetic field ($B_1$) \cite{cryst5010014}, while the operator $\mat{X}$ transforms like a vector (it is one partner of an irreducible 2D representation $E$).
The normal incidence Maxwell operator $\mat{M}_0$ thus transforms like $A$, and $\mat{M}_1$ transforms like $E$.
This induces \emph{transverse} fields which transform like $E$ for $\rm{d}\theta\inl\ne0$ (see detailed perturbation theory below).
To avoid using the 2D irreducible representation $E$, where the $C_4$ rotations and $\sigma_d$ mirrors couple $\rm{d}k_x$ with $\rm{d}k_y$ perturbations, we classify the transverse fields according to the subgroup $C_{2v}$, shown in Table \ref{tab:C4v}.
The $E$ representation in $C_{4v}$ reduces to the $B_2$ and $B_3$ 1D representations in $C_{2v}$ \cite{Dresselhaus}.
The symmetry behaviour of $\mat{M}_0$ and $\mat{M}_1$ in $C_{2v}$ reveals that the transverse electric fields must transform like $B_3$ and the magnetic fields like $B_2$ in $C_{2v}$.
The symmetry behaviour of all operators and modal fields is given in Table \ref{tab:C4v}.

\begin{table}
    \centering
    \begin{tabular}{r|ccccc|c}
        $C_{4v}$ & $E$ & $C_2$ & $2C_4$ & $2\sigma$ & $2\sigma_d$ & fields/oper.  \\\hline
        $A$      & 1   &  1    &  1     &  1        &  1          & $\El$, $\mat{M}_0$       \\
        $B_1$    & 1   &  1    &  1     & -1        & -1          & $\Hl$, curl \\
        $B_2$    & 1   &  1    & -1     &  1        & -1          & \\
        $B_3$    & 1   &  1    & -1     & -1        &  1          & \\
        \multirow{1}{*}{$E$}      & \multirow{1}{*}{2}   & \multirow{1}{*}{-2}    &  \multirow{1}{*}{0}     &  \multirow{1}{*}{0}        &  \multirow{1}{*}{0}          & \multirow{1}{8em}{$\Et$, $\Ht$, $\mat{X}$, $\mat{M}_1$}\\
    \end{tabular}\hfill
    \begin{tabular}{r|cccc|c}
        $C_{2v}$ & $E$ & $C_2$ & $\sigma_x$ & $\sigma_y$ & fields/oper.  \\\hline
        $A$      & 1   &  1    &  1         &  1         & $\El$, $\mat{M}_0$       \\
        $B_1$    & 1   &  1    & -1         & -1         & $\Hl$, curl \\
        $B_2$    & 1   & -1    &  1         & -1         & $\Ht$, $\mat{X}$  \\
        $B_3$    & 1   & -1    & -1         &  1         & $\Et$, $\mat{M}_1$
    \end{tabular}
    \vspace{1em}
    \caption{Character table for the point group $C_{4v}$ and its subgroup $C_{2v}$ containing the identity $E$, a \SI{180}{\degree} rotation $C_2$, a $\inl\pm\SI{90}{\degree}$ rotation $C_4$, horizontal ($\sigma_x$) and vertical ($\sigma_y$) mirrors $\sigma$ and two diagonal mirrors $\sigma_d$.}
    \label{tab:C4v}
\end{table}

\subsection{Expansion of the fields at the PDN surface}\label{sec:supp:expansion}
We can now expand the $x$-$y$ dependence of the fields in ascending Bragg order with the required point symmetry behaviour.
We introduce the notation $\vec{F}_{sn}^{\{hk\}}$, where $s\eq \rm{l},\rm{t}$ denotes the point symmetry behaviour (\emph{longitudinal} or \emph{transverse}), $\{hk\}$ with $h\inl\geq k\geq 0$ the Bragg order\footnote{The field is a linear combination of all in-plane reciprocal lattice vectors of equal length, which is an equivalence class as defined in \cite{saba2013}. It is generated by applying the $C_{4v}$ point group to the representative vector with Miller index in the lower half of the upper right quadrant of $\mathbb{N}^2$.}, and $n$ is an additional integer label to distinguish between different fields for a given $s$ and $\{hk\}$ combination.

Let us first concentrate on the longitudinal basis fields.
For $h\inl>k\inl>0$ we obtain six basis fields -- three electric and three magnetic fields. Introducing the short-hand notation $(X,Y)\deq 2\pi/a(x,y)$, these are
\begin{subequations}
\begin{align}
    \vec{F}_{l1}^{\{hk\}} &= \begin{pmatrix}
        \sin(hX)\cos(kY) \\
        \sin(hY)\cos(kX) \\ 0 \\ 0 \\ 0 \\ 0
    \end{pmatrix}\;\text{;}\quad
    \vec{F}_{l2}^{\{hk\}} = \begin{pmatrix}
        \sin(kX)\cos(hY) \\
        \sin(kY)\cos(hX) \\ 0 \\ 0 \\ 0 \\ 0
    \end{pmatrix}\;;\notag\\
    \vec{F}_{l3}^{\{hk\}} &= \begin{pmatrix}
        0 \\ 0 \\ \cos(hX)\cos(kY) + \cos(kX)\cos(hY) \\ 0 \\ 0 \\ 0
    \end{pmatrix}\,;\\
    \vec{F}_{l4}^{\{hk\}} &= \begin{pmatrix}
        0 \\ 0 \\ 0 \\
        -\sin(hY)\cos(kX) \\ \phantom{-}\sin(hX)\cos(kY) \\  0
    \end{pmatrix}\;;\quad
    \vec{F}_{l5}^{\{hk\}} = \begin{pmatrix}
        0 \\ 0 \\ 0 \\
        -\sin(kY)\cos(hX) \\ \phantom{-}\sin(kX)\cos(hY) \\  0
    \end{pmatrix}\;;\notag\\
    \vec{F}_{l6}^{\{hk\}} &= \begin{pmatrix}
        0 \\ 0 \\ 0 \\ 0 \\ 0 \\ \sin(hX)\sin(kY) - \sin(kX)\sin(hY)
    \end{pmatrix}\;.
\end{align}
\end{subequations}
For $h\eq k\inl>0$, $\vec{F}_{l1}^{\{hk\}}\eq\vec{F}_{l2}^{\{hk\}}$ and $\vec{F}_{l4}^{\{hk\}}\eq\vec{F}_{l5}^{\{hk\}}$, while $\vec{F}_{l6}^{\{hk\}}\eq0$, leaving one independent magnetic field and two electric fields.
For $k\eq0$ and $h\inl\neq0$, $\vec{F}_{l2}^{\{hk\}}\eq\vec{F}_{l5}^{\{hk\}}\eq\vec{F}_{l6}^{\{hk\}}\eq0$, also leaving one independent magnetic field and two electric fields.
For $h\eq0$, only $\vec{F}_{l3}^{\{hk\}}$ is non-zero, which is why the field in the homogenised picture is a purely longitudinal electric field.

For the transverse basis, the vector components of the field decouple and the Bragg orders are classified by $(hk)\inl\in \mathbb{N}_+^2$. For general $h\inl\neq k$ we obtain three electric and three magnetic basis fields
\begin{subequations}
\begin{align}
    \vec{F}_{t1}^{\{hk\}} &= \begin{pmatrix}
         \cos(hX)\cos(kY) \\ 0 \\ 0 \\ 0 \\ 0 \\ 0
    \end{pmatrix}
    \vec{F}_{t2}^{\{hk\}} = \begin{pmatrix}
         0 \\ \sin(hX)\sin(kY) \\ 0 \\ 0 \\ 0 \\ 0
    \end{pmatrix}
    \vec{F}_{t3}^{\{hk\}} = \begin{pmatrix}
         0 \\ 0 \\ \sin(hX)\cos(kY) \\ 0 \\ 0 \\ 0
    \end{pmatrix} \,\text{;}\label{eq:transfields1}\\
    \vec{F}_{t4}^{\{hk\}} &= \begin{pmatrix}
         0 \\ 0 \\ 0 \\ \sin(hX)\sin(kY) \\ 0 \\ 0
    \end{pmatrix}
    \vec{F}_{t5}^{\{hk\}} = \begin{pmatrix}
         0 \\ 0 \\ 0 \\ 0 \\ \cos(hX)\cos(kY) \\ 0
    \end{pmatrix}
    \vec{F}_{t6}^{\{hk\}} = \begin{pmatrix}
         0 \\ 0 \\ 0 \\ 0 \\ 0 \\ \cos(hX)\sin(kY)
    \end{pmatrix} \,\text{.}\label{eq:transfields2}
\end{align}
\label{eq:transfields}
\end{subequations}
As for the longitudinal case, some of these fields disappear when $h\eq k$ or when one or two indices are zero.
Table \ref{tab:surf_basis} lists the field basis for the lowest Bragg orders considered.
These basis fields are conveniently orthonormal in the canonical inner product on the 2D surface unit cell (UC),
\begin{equation}
    \langle F,G\rangle := \frac{1}{V_{\rm{UC}}} \int_{\rm{UC}} \rm{d}x\rm{d}y\;\vec{F}^*\cdot\vec{G} \,,
    \label{eq:inner}
\end{equation}
which is
\begin{equation}
    \langle \vec{F}_{sn}^{\{hk\}},\vec{F}_{\tilde{s}\tilde{n}}^{\{\tilde{h}\tilde{k}\}}\rangle := \delta_{s\tilde{s}}\delta_{n\tilde{n}}\delta_{h\tilde{h}}\delta_{k\tilde{k}} \,.
    \label{eq:perp}
\end{equation}

\begin{table}
    \centering
    \begin{tabular}{r|cccccc}
         $s$        &   l   &   l   &   l   &   l   &   t   &   t   \\
         $\{hk\}$   &   00  &   10  &   10  &   10  &   00  &   00  \\
         $n$        &   1   &   1   &   2   &   3   &   1   &   2   \\\hline
         $\vec{F}$  &   $\begin{pmatrix}  0 \\ 0 \\ 1 \\ 0 \\ 0 \\ 0 \end{pmatrix}$ &
                        $\begin{pmatrix} \sin(bx) \\ \sin(by) \\ 0 \\ 0 \\ 0 \\ 0 \end{pmatrix}$ &
                        $\begin{pmatrix} 0 \\ 0 \\ \cos(bx)+\cos(by) \\ 0 \\ 0 \\ 0 \end{pmatrix}$ &
                        $\begin{pmatrix} 0 \\ 0 \\ 0 \\ {-}\sin(by) \\ \sin(bx) \\ 0 \end{pmatrix}$ &
                        $\begin{pmatrix} 1 \\ 0 \\ 0 \\ 0 \\ 0 \\ 0 \end{pmatrix}$ &
                        $\begin{pmatrix} 0 \\ 0 \\ 0 \\ 0 \\ 1 \\ 0 \end{pmatrix}$
    \end{tabular}\vspace{1em}
    \caption{Surface field basis to express the electromagnetic modes at the slab interface. We consider longitudinal fields up to the first Bragg order and transverse fields in the zero Bragg order ($b\deq 2\pi/a$ is the reciprocal lattice constant).}
    \label{tab:surf_basis}
\end{table}

\subsection{Surface-normal BIC solutions for the PDN slab}\label{sec:supp:BICfreqs}
The vacuum modes in the semi-infinite space above the slab ($z\inl>z_0$) can be calculated analytically.
Due to translation symmetry, the solution in free space is of the Rayleigh form \cite{cryst5010014}, which is a superposition of modal solutions within the individual Bragg orders.
For normal radiation \eqref{eq:MWE_modal} has only the trivial solution in the zero Bragg order as $\mat{M}_0 \vec{F}_{l1}^{\{00\}}\eq{-}\vec{F}_{l1}^{\{00\}}$, i.e.\ the zero Bragg order field is not in the nullspace of $\mat{M}_0$.

Consider the first Bragg order, expressed by
\begin{equation}
    \vec{F}_{\rm{V}l}^{\{10\}} = \sum_n \vec{F}_{ln}^{\{10\}} c_n \,,
    \label{eq:B1l}
\end{equation}
with a complex coefficient vector $\vec{c}\deq(c_1,c_2,c_3)$. 
In vacuum, the operator $\mat{M}_0$ becomes
\begin{equation}
    \left(\mat{M}_0\right)_{\rm{l}\rm{l}}^{\{10\}} =
    \begin{pmatrix}
        -1 & 0 & k_z/k_0 \\
        0 & -1 & \imath b/k_0 \\
        k_z/k_0 & -\imath b/k_0 & -1
    \end{pmatrix}
    \label{eq:M10ll}
\end{equation}
when acting on the symmetry basis vectors $\vec{F}_{ln}^{\{10\}}$.
For $\rm{d}\theta\eq0$, \eqref{eq:MWE_modal} is thus solved by the (normalised) coefficient eigenvector
\begin{equation}
    \vec{c}_l^{\{10\}} = \frac{1}{\sqrt{2}b} \begin{pmatrix}
                            \kappa_{\{10\}} \\ b \\ -\imath k_0
                        \end{pmatrix}\,,
     \label{eq:B1coeff_disp}
\end{equation}
with the attenuation constant $\kappa_{\{10\}} \deq {-}\imath k_z\eq\sqrt{b^2-k_0^2}$ satisfying the vacuum dispersion.

Within the PDN slab, the modal solutions of the EAW wave are determined by the counter-propagating currents in the two networks \cite{Wang2023}.
Since we are interested in the termination behaviour, including the tennis net region, the thin wire approximation \cite{Silve2005} is not quite appropriate.
In the former, the wires carry a current dominated by the Bloch wave dispersion along $z$ for small frequencies, with a small additional sawtooth term that causes the currents to alternately charge the tennis net grids.
For finite size square shaped rods, instead, we find that each rod carries a current in the $z$ direction that mainly charges the top and bottom plates of the tennis nets attached to its bottom and top.
The associated charge profile has the characteristics of a line dipole giving rise to a negligible electric field in the zero Bragg order in the tennis net region and a field that varies approximately according to the Bloch dispersion in the rod region.

These general observations are well reproduced by full-wave simulations, from which we project the relevant field components.
First, the fields are expanded into the surface in-plane basis functions
\begin{equation}
    \vec{F}_{\rm{PDN}} = \sum_{h\geq k}\sum_n \vec{F}_{ln}^{\{hk\}}(x,y) g_{ln}^{\{hk\}}(z)
    \label{eq:MMmodes}
\end{equation}
with periodic functions $g_{ln}^{\{hk\}}(z)$.
Using \eqref{eq:perp} we thus calculate the $g_{ln}^{\{hk\}}(z_{\tau})$ at the termination plane $z\eq z_{\tau}$
\begin{equation}
    g_{ln}^{\{hk\}}(z_{\tau}) = \left\langle \vec{F}_{ln}^{\{hk\}} ,\vec{F}_{\rm{PDN}}(z_{\tau}) \right\rangle  \,.
    \label{eq:projection}
\end{equation}

To gain a basic understanding within a simple analytical model, we restrict ourselves to the dominant fundamental bulk modes within the PDN slab: In the case of the tennis net domain, only the EAW wave propagating energy through the slab is sufficient.
For the rod domain, where the zero Bragg order longitudinal electric field is finite, we add the evanescent wave with the lowest numerical attenuation $\kappa_{\rm{eva}}\eq 1.03 2\pi/a$ to match the number of degrees of freedom with the number of boundary conditions.

We use an additional symmetry classification to restrict the scattering problem to the top surface.
The tennis net termination with even $N$ is mirror symmetric with respect to the centre of the slab (yellow mirror plane in \figref{fig:nets-and-resonances} b), with mirror symmetry $\mathcal{M}$.
This mirror symmetry transforms the EAW wave into the counter-propagating wave with negated $z$-components
\begin{align}
    \mathcal{M}\,\vec{F}(\vec{r})e^{\imath kz} &= \begin{pmatrix}
        \phantom{-}E_x(x,y,-z) \\ \phantom{-}E_y(x,y,-z) \\ -E_z(x,y,-z) \\
        -H_x(x,y,-z) \\ -H_y(x,y,-z) \\ \phantom{-}H_z(x,y,-z)
    \end{pmatrix}e^{-\imath kz}\,.
    \label{eq:mirrorsymm}
\end{align}
Note that we have to treat the magnetic field as a pseudo-vector due to the ${-}1$ character of the curl in Maxwell's equations regarding improper rotations.
The time reversal symmetry $\mathcal{T}$ also transforms the EAW wave into the counter-propagating wave, but with complex conjugated field components instead
\begin{align}
    \mathcal{T}\,\vec{F}(\vec{r})e^{\imath kz} &= \begin{pmatrix}
        \phantom{-}E_x^*(\vec{r}) \\ \phantom{-}E_y^*(\vec{r}) \\ \phantom{-}E_z^*(\vec{r}) \\
        -H_x^*(\vec{r}) \\ -H_y^*(\vec{r}) \\ -H_z^*(\vec{r})
    \end{pmatrix}e^{-\imath kz}\,.
    \label{eq:TRsymm}
\end{align}
The magnetic field also acquires an additional negative sign due to the ${-}1$ character of the time derivative in Maxwell's equations regarding time reversal.
The mirror-time symmetry $\mathcal{M}\mathcal{T}$ thus maps the non-degenerate EAW wave onto itself, leading to
\begin{align*}
    \mathcal{M}\mathcal{T}\,\vec{F}(\vec{r})e^{\imath kz} &= \begin{pmatrix}
        \phantom{-}E_x^*(x,y,-z) \\ \phantom{-}E_y^*(x,y,-z) \\ -E_z^*(x,y,-z) \\
        \phantom{-}H_x^*(x,y,-z) \\ \phantom{-}H_y^*(x,y,-z) \\ -H_z^*(x,y,-z)
    \end{pmatrix}e^{\imath kz}\,.
\end{align*}
Therefore, we can normalise the fields (gauge freedom) so that all in-plane fields $g_{sn}^{\{hk\}}\inl\in\mathbb{R}$ and all out-of-plane fields $g_{sn}^{\{hk\}}\inl\in\imath\mathbb{R}$;
and the downward and upward propagating waves are related by \eqref{eq:mirrorsymm}.\footnote{
There is an additional $\pm$ phase choice depending on whether the slab mode is even or odd with respect to the mirror symmetry.
However, the first excited mode we are interested in has one node in the $z$ component of the current.
Therefore, the electric field is even, so that the sign choice in \eqref{eq:mirrorsymm} and \eqref{eq:TRsymm} is correct to obtain the counter-propagating wave.}
The field at the top of the PDN is thus approximated by the downward propagating EAW wave and the upward propagating wave, which is a symmetry copy according to \eqref{eq:mirrorsymm}.
However, the latter acquires a phase $\pm p\eq\exp\{\imath k_0 N a/\kappa\}$ as it travels through the slab to the top surface.
The same is true for the evanescent mode, but the phase is not on the complex unit circle but a small real number that corresponds to a more than 20-fold decrease in the field intensity through the slab.
We, therefore, only need to consider the evanescent field at the top surface in good approximation.

In the vacuum domain, the field is approximated by the analytical solution \eqref{eq:B1l}.
Since the $E_z$ field vanishes to a good approximation in the tennis net domain, as discussed above, we require $\vec{F}_{\rm{l}1}^{\{10\}}$ and $\vec{F}_{\rm{l}3}^{\{10\}}$ to be continuous at the top interface, yielding the following $2\inl\times2$ generalised eigenproblem for the phase $p$ as eigenvalue
and the free coefficients $c_{\rm{d}}$ and $c_{\rm{V}}$ of the EAW wave and the first Bragg order vacuum mode as the eigenvector
\begin{align}
    \begin{pmatrix}
        1 & 1 \\ -G & -\imath k_0/\kappa_{\{10\}}
    \end{pmatrix} \begin{pmatrix}
        c_{\rm{d}} \\ c_{\rm{V}}
    \end{pmatrix} = - p
    \begin{pmatrix}
        \phantom{-}1 & 0 \\ G & 0
    \end{pmatrix} \begin{pmatrix}
        c_{\rm{d}} \\ c_{\rm{V}}
    \end{pmatrix}\,.
    \label{eq:TNeigen}
\end{align}
We have conveniently normalised the coefficient vectors of the bulk modes so that their electric field coefficient is one.
The positive number $G\deq {-}g_{\rm{l}3}^{\{hk\}}(z_{\tn})/g_{\rm{l}1}^{\{hk\}}(z_{\tn})$, which is numerically close to $G\eq\num{0.8}$, determines the magnetic field coefficient of the EAW.
The magnetic coefficient of the vacuum field becomes $-\imath k_0/\kappa_{\{10\}}$ according to \eqref{eq:B1coeff_disp}, which is well approximated by $\delta\deq k_0/\kappa_{\{10\}}\,{\approx}\,0.08$ at the first BIC resonance with the hard wall approximation \eqref{eq:FSR}, which assumes $p\eq{-}1\eq\exp\{\imath\pi\}$.
The eigenproblem is solved when
\begin{equation}
    p = -\frac{1-\imath\delta/G}{1+\imath\delta/G} = e^{\imath[\pi-2\arctan(\delta/G)]} \approx e^{\imath(\pi-0.2)}\,.
\end{equation}
This corrects the phase and therefore the BIC frequency.
It is $\nu_{\rm{BIC}}\,{\approx}\,\SI{8.75}{\giga\hertz}$ for the hard wall boundary condition in the homogenised picture according to \eqref{eq:FSR} with $\kappa\,{\approx}\,0.7$ for the square rod PDN discussed in the main text.
The phase-corrected theoretical frequency is $\nu_{\rm{BIC}}(z_{\tn})\eq(1-0.2/\pi)\nu_{\rm{BIC}}\,{\approx}\,\SI{8.19}{\giga\hertz}$ in good agreement with the simulation results (see Table \ref{tab:Q_F}).

For the rods termination, the zero Bragg order electric field cannot be ignored.
To obtain a well-defined problem, this requires adding the first evanescent Floquet mode inside the PDN slab, which has a numerical attenuation constant of $\kappa\,{\approx}\,1.025 b$, yielding a corresponding $3{\times}3$ eigenproblem.\footnote{The attenuation close to the reciprocal lattice constant $b$ makes sense as the field is close to the first Bragg order vacuum field in the rods domain, which approximately satisfies the PEC boundary condition at the rod positions.}
Additionally, the slab with an even number of unit cells does not feature a mirror symmetry in its centre, but a glide-mirror plane that additionally shifts the slab by half a unit cell $a/2$ in both $x$ and $y$ directions.
Therefore, when relating the upwards and downwards propagating EAW waves via \eqref{eq:mirrorsymm}, any odd Bragg order with $h{+}k\,{\in}\,2\mathbb{N}{+}1$ acquires an additional minus sign in all components.
Solving the eigenproblem yields a corrected frequency of \SI{8.57}{\giga\hertz} that is closer to the hard-wall boundary condition (as it contains the zero Bragg order) and close to the simulation result (see Table \ref{tab:Q_F}).

\subsection{Radiating quasi-BIC solutions for small radiation angle}\label{sec:supp:quasiBIC}

To obtain the bulk modes for a small radiation angle, we return to $\mat{M}_0\vec{F}\eq\rm{d}\theta\,\mat{M}_1\vec{F}$ (\eqref{eq:MWE_modal}) for a small but finite angle $\rm{d}\theta$.
Since $\mat{M}_0$ leaves the symmetry channels invariant and $\mat{M}_1$ generates a transverse contribution from the longitudinal field, this equation induces a cross-coupling between the two channels.
We restrict the basis of the transverse channel to the zero Bragg order, which describes the radiating wave in vacuum, see $F_{\rm{t}1}^{\{00\}}$ and $F_{\rm{t}2}^{\{00\}}$ in Table \ref{tab:surf_basis}.
At normal incidence, it is coupled to an evanescent transverse wave in the PDN slab.
It has a numerical attenuation constant of $\kappa\,{\approx}\,0.572 b$, and projected field components depending on the termination.
Since there is no propagating wave inside the slab in the transverse channel, there is no transverse solution below the effective PDN plasma frequency at normal incidence.

The perturbed eigenproblem at non-normal radiation reads
\begin{align}
    \begin{pmatrix}
        \mat{A}_{\rm{ll}} & \mat{A}_{\rm{lt}}\rm{d}\theta \\
        \mat{A}_{\rm{tl}}\rm{d}\theta & \mat{A}_{\rm{tt}}
    \end{pmatrix} \begin{pmatrix}
        \vec{c}_{\rm{l}} \\ \vec{c}_{\rm{t}}
    \end{pmatrix} = (p_0 + \rm{d}p)
    \begin{pmatrix}
        \mat{B}_{\rm{ll}} & 0 \\
        \mat{B}_{\rm{tl}}\rm{d}\theta & 0
    \end{pmatrix} \begin{pmatrix}
        \vec{c}_{\rm{l}} \\ \vec{c}_{\rm{t}}
    \end{pmatrix}
    \,,
    \label{eq:NonNormalEigen}
\end{align}
where $A_{\rm{ll}}\,{\in}\,\rm{GL}_3(\mathbb{C})$ contains the normal incidence projected fields of the three longitudinal modes (EAW, evanescent in material, and first Bragg order vacuum; columns) onto the three surface fields (rows).
$A_{\rm{tt}}\,{\in}\,\rm{GL}_2(\mathbb{C})$ contains the two transverse field components (evanescent in material and propagating zero Bragg order in vacuum), projected onto the two transverse surface fields.
$B_{\rm{ll}}$ contains the components of the counter-propagating EAW (as discussed above) and vanished in the other columns, \textit{cf.} the RHS of \eqref{eq:TNeigen}.
To obtain the cross-coupling matrices, we solve \eqref{eq:MWE_modal} for the individual modes under two assumptions: First, the action of the curls in $\mat{M}_0$ is dominated by the vacuum region even in the PDN slab.
This is a good approximation, particularly in the rods region, since the modes already satisfy the PEC boundary conditions and the volume of the struts is small.
Second, the modulation of the Floquet waves with $z$ is dominated by the Bloch phase $\exp\{\imath k_z z\}$.
That is, the periodic function $g(z)$ is flat at the termination plane.
This is, again, only a good approximation in the centre of the rods domain, where the field is mainly generated by the rod currents that carry the Bloch phase advance.

The elements of the cross-coupling matrix $A_{\rm{tl}}$ are then obtained by solving \eqref{eq:MWE_modal}
\begin{equation*}
    \left((\mat{M}_{0})_{\rm{tt}}^{\{00\}}\right)^{-1}\,(\mat{M}_{1})_{\rm{tl}}^{\{00\}} g_{\rm{l}2}^{\{00\}}\,,
\end{equation*}
where the field corresponds to the projections for the different modes (EAW, evanescent, first Bragg order) in the different columns, and the matrices are the operators in the zero Bragg order bases with $k_z$ of the corresponding mode
\begin{equation*}
    (\mat{M}_{0})_{\rm{tt}}^{\{00\}} = \begin{pmatrix}
        -1 & \frac{k_z}{k_0} \\ \frac{k_z}{k_0} & -1
    \end{pmatrix}   \quad\text{and}\quad
    (\mat{M}_{1})_{\rm{tl}}^{\{00\}} = \begin{pmatrix}
        0 \\ 1
    \end{pmatrix}\,.
\end{equation*}
For $B_{\rm{tl}}$ we apply the same equation to the upwards propagating EAW wave's coefficients.
The row in $A_{\rm{lt}}$ that corresponds to $F_{\rm{l}2}^{\{00\}}$ is obtained by solving
\begin{equation*}
    \left((\mat{M}_{0})_{\rm{ll}}^{\{00\}}\right)^{-1}\,(\mat{M}_{1})_{\rm{lt}}^{\{00\}}
    \begin{pmatrix}
          g_{\rm{t}1}^{\{00\}}  \\
          g_{\rm{t}2}^{\{00\}} 
    \end{pmatrix}
\end{equation*}
for each of the two modes (evanescent and zero Bragg order), with
\begin{equation*}
    (\mat{M}_{0})_{\rm{ll}}^{\{00\}} = \begin{pmatrix}
        -1
    \end{pmatrix}   \quad\text{and}\quad
    (\mat{M}_{1})_{\rm{lt}}^{\{00\}} = \begin{pmatrix}
        0 & 1
    \end{pmatrix}\,.
\end{equation*}
The rows in $A_{\rm{lt}}$ that correspond to $F_{\rm{l}1}^{\{10\}}$ and $F_{\rm{l}2}^{\{10\}}$ are obtained by solving
\begin{equation*}
    \left((\mat{M}_{0})_{\rm{ll}}^{\{10\}}\right)^{-1}\,(\mat{M}_{1})_{\rm{lt}}^{\{10\}}
    \begin{pmatrix}
          g_{\rm{t}3}^{\{10\}}  \\
          g_{\rm{t}5}^{\{10\}}+c_{\rm{t}5}^{\{01\}}  \\
          g_{\rm{t}6}^{\{10\}}
    \end{pmatrix}\,,
\end{equation*}
where the vector contains the field projections onto the only transverse basis fields in the first Bragg order, see \eqref{eq:transfields}, that are non-zero and that $\mat{M}_1$ maps onto the longitudinal channel (trivial representation in Table \ref{tab:C4v}).
The matrices are, \textit{cf.} \eqref{eq:M10ll}
\begin{equation*}
    (\mat{M}_{0})_{\rm{ll}}^{\{10\}} = 
    \begin{pmatrix}
        -1 & 0 & k_z/k_0 \\
        0 & -1 & \imath b/k_0 \\
        k_z/k_0 & -\imath b/k_0 & -1
    \end{pmatrix}   \quad\text{and}\quad
    (\mat{M}_{1})_{\rm{lt}}^{\{10\}} = \begin{pmatrix}
        0 & 0 & -\frac{1}{2} \\ 0 & 1 & 0 \\ \frac{1}{2} & 0 & 0
    \end{pmatrix}\,.
\end{equation*}

We have now populated the matrices in the scattering eigenproblem \eqref{eq:NonNormalEigen} and can solve it perturbatively.
\eqref{eq:NonNormalEigen} thus yields the two equations
\begin{subequations}
\begin{align}
    (\mat{A}_{\rm{ll}}-p_0\mat{B}_{\rm{ll}})\vec{c}_{\rm{l}} + \rm{d}\theta\,\mat{A}_{\rm{lt}}\vec{c}_{\rm{t}} &= \rm{d}p\,\mat{B}_{\rm{ll}}\vec{c}_{\rm{l}} \label{eq:NonNormalEigen1} \\
    \quad \text{and} \quad
    \mat{A}_{\rm{tt}}\vec{c}_{\rm{t}} 
    &= \rm{d}\theta\,\left(p_0 \mat{B}_{\rm{tl}}-\mat{A}_{\rm{tl}}\right)\vec{c}_{\rm{l}}\,. \label{eq:NonNormalEigen2}
\end{align}
\end{subequations}
For small $\rm{d}\theta$, we can test \eqref{eq:NonNormalEigen1} with the left eigenvector $\vec{u}_0$ of the unperturbed problem that solves
\begin{align*}
    (\mat{A}_{\rm{ll}}-p_0\mat{B}_{\rm{ll}})^{\phantom{\dagger}}\, \vec{v}_0 &= 0 \\
    (\mat{A}_{\rm{ll}}-p_0\mat{B}_{\rm{ll}})^\dagger\, \vec{u}_0 &= 0
\end{align*}
so that the equation becomes an algebraic equation without $p_0$
\begin{equation}
    \rm{d}\theta\,\langle \vec{u}_0,\mat{A}_{\rm{lt}}\vec{c}_{\rm{t}}\rangle = \rm{d}p\,\langle \vec{u}_0,\mat{B}_{\rm{ll}}\vec{c}_{\rm{l}}\rangle \label{eq:NonNormalEigen1tested}
\end{equation}
Since we projected the original problem out, we can now safely approximate $\vec{c}_{\rm{l}}\eq\vec{v}_0$ in lowest order.
Since $\mat{A}_{\rm{tt}}$ is invertible in \eqref{eq:NonNormalEigen2}, we can further eliminate $\vec{c}_{\rm{t}}$ in \eqref{eq:NonNormalEigen1tested} to obtain
\begin{align}
    \rm{d}p = \underbrace{\frac{\langle \vec{u}_0,\mat{A}_{\rm{lt}}\mat{A}_{\rm{tt}}^{-1}\left(p_0 \mat{B}_{\rm{tl}}-\mat{A}_{\rm{tl}}\right)\vec{v}_0\rangle}{\langle \vec{u}_0,\mat{B}_{\rm{ll}}\vec{v}_0\rangle}}_{p_1}\,\rm{d}\theta^2\,\,.
\end{align}
We, therefore, obtain the frequency shift at finite $\rm{d}\theta$ using
\begin{equation}
    p(\rm{d}\theta)\approx p_0+p_1\,\rm{d}\theta^2 = e^{\imath \alpha(\rm{d}\theta) (f_0+f_1\rm{d}\theta^2)}\,,
\end{equation}
where $f_0$ is the BIC frequency at the $\Gamma$ point and
\begin{equation}
\alpha(\rm{d}\theta) = \underbrace{\frac{2\pi N a}{c\kappa}}_{=-\imath \ln(p_0)/f_0}\cos(\rm{d}\theta)\approx-\imath\, \frac{\ln(p_0)}{f_0}(1-\rm{d}\theta^2/2)\,.
\end{equation}
In the second order, we, therefore, obtain
\begin{equation}
    f_1 \approx \left(\frac{1}{2}+\frac{p_1}{p_0\ln(p_0)}\right) f_0 \quad\text{and}
    \quad Q_{\rm{freq}} = \frac{f_0}{\Im f_1}\,\rm{d}\theta^{-2}\approx \frac{1}{\Im\{\frac{p_1}{p_0\ln(p_0)}\}}\,\rm{d}\theta^{-2}\,.
\end{equation}
This yields $Q_{\rm{freq}}(\rods)\,{\approx}\,\num{1.0e3}\rm{d}\theta^{-2}$ for the rods termination and $Q_{\rm{freq}}(\tn)\,{\approx}\,\num{5.1e3}\rm{d}\theta^{-2}$ for the tennis net termination.
As expected, the prediction for the rods termination is close to the simulation results at $\theta\eq\pi/16$, while it underestimates the simulated Q factor for the tennis net termination significantly.

Since the imaginary part of the phase correction can be almost tangential to the complex unit circle, the frequency-based Q factors become numerically unstable, particularly for the tennis net termination, for which some of the approximations made are less well justified.
We, therefore, go back to the definition of the Q factor from resonator theory, namely $2\pi$ times the quotient of the stored energy in the resonator divided by the energy loss per optical cycle \cite{Saleh2019}.
Most of the stored energy is in the two counter-propagating EAW waves in the PDN slab.
The energy per surface area is then in the homogenisation approximation given by $2Nau_{\rm{EAW}}\,{\approx}\,8a\varepsilon_0|E_{\rm{EAW}}|^2$, where we used $N\eq4$ and the electro-magnetic energy density of the longitudinal EAW plane wave $u_{\rm{EAW}}\eq n^2\varepsilon_0/2|E_{\rm{EAW}}|^2$ with an effective refractive index $n^2\eq1/\kappa^2\,{\approx}\,2$ and the electric field $E_{\rm{EAW}}$ of the EAW wave.
The radiated intensity on both sides is $2I_{B_0}\eq c\varepsilon_0|E_{B_0}|^2$, while the optical cycle is $2\pi/\Re\{\omega\}$.
After normalising the fields in the surface basis under the assumption that the surface-averaged EAW energy is constant throughout the unit cell, this yields
\begin{equation}
    Q_{\rm{field}} = 8\Re\{k_0\}a\frac{|c_{\rm{EAW}}|^2}{|c_{B_0}|^2}\,.
\end{equation}
The field quality factors thus evaluate to $Q_{\rm{field}}(\rods)\,{\approx}\,\num{5.4e2}\rm{d}\theta^{-2}$ for the rods termination and $Q_{\rm{freq}}(\tn)\,{\approx}\,\num{1.2e6}\rm{d}\theta^{-2}$ for the tennis net termination.
If the physical resonator is well described, the field and the frequency formulation of the quality factor should, of course, be identical (up to a factor of two) due to energy conservation.
For the rods termination, the two expressions are indeed close to each other and predict the simulation result at $\theta\eq\pi/16$ well, see Table \ref{tab:Q_F}.
For the tennis net termination, they differ by more than two orders of magnitude mainly because the wave impedance is not well approximated by the Bloch dispersion.
Nevertheless, the logarithmic average of the field and the frequency Q factors predicts $\ln(Q)\eq 6.3$ for $\theta\eq\pi/16$, very close to the simulation result of $\ln(Q)\eq6.5$.

\newpage
\renewcommand{\thesection}{S2}
\section{Additional Figures}

\renewcommand{\thefigure}{S1}
\begin{figure}[h]
\includegraphics[width=\textwidth]{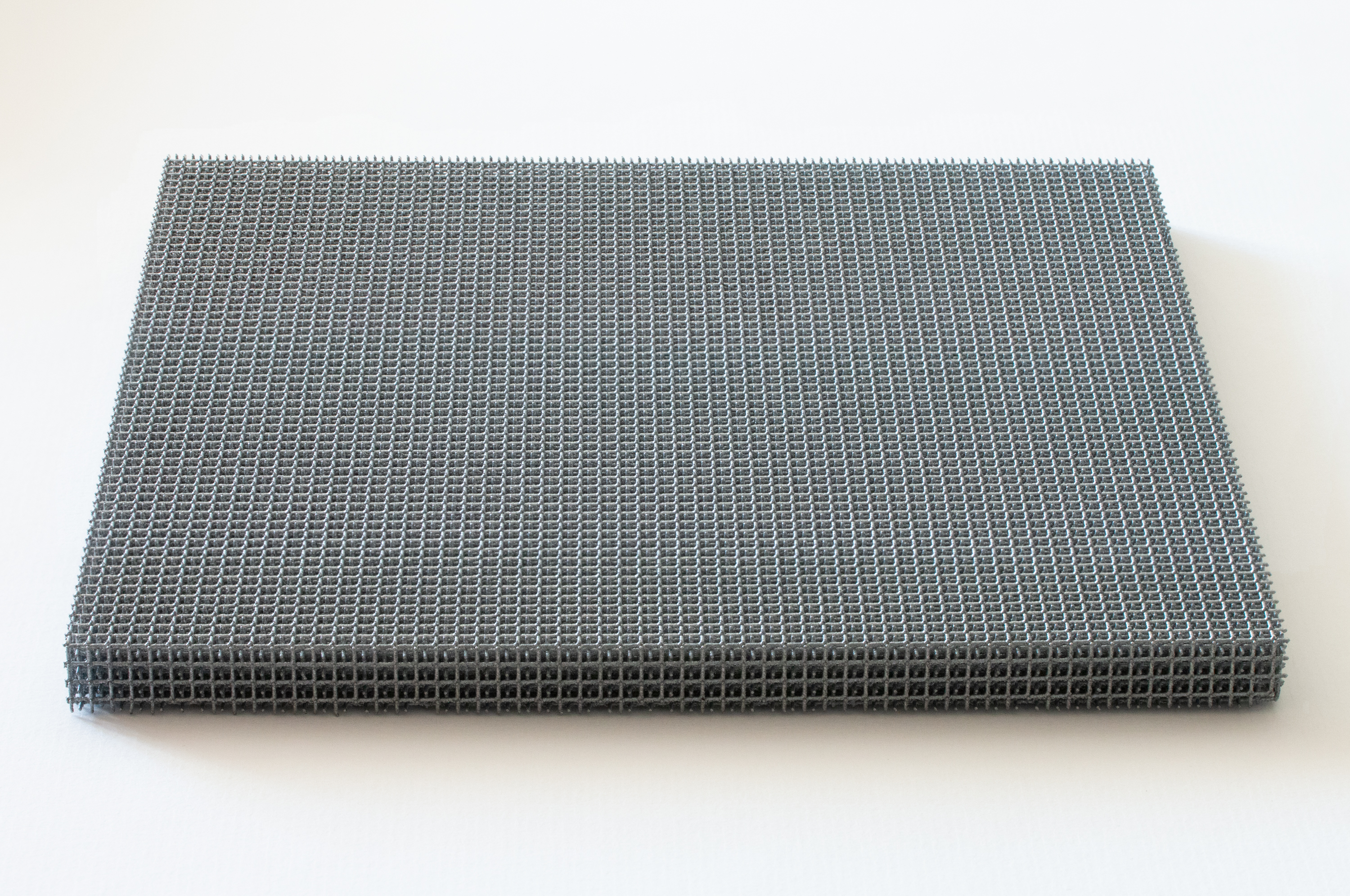}
\centering
\caption{View of the full 60x60x4 unit cell sample. The cubic unit cell size is 3 mm yielding a total sample size of 180x180x12 mm. The image was stacked to improve the depth of field.}
\label{fig:supp:full_sample}
\end{figure}

\newpage
\renewcommand{\thefigure}{S2}
\begin{figure}[h]
\includegraphics[width=\textwidth]{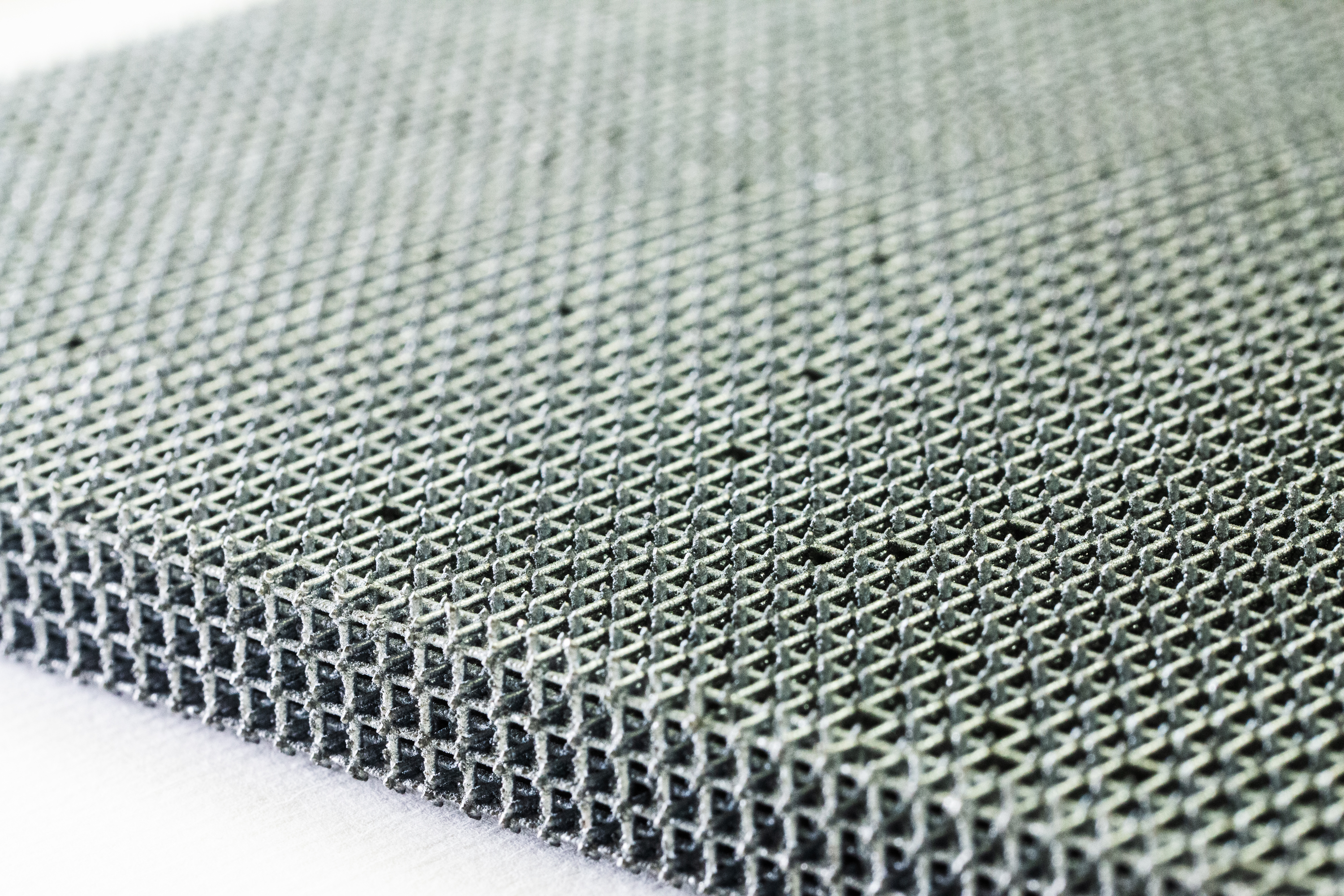}
\centering
\caption{Macro photography of the sample back showing a few missing struts. This defect originated while the print supports were removed.}
\label{fig:supp:defects}
\end{figure}

\newpage
\renewcommand{\thefigure}{S3}
\begin{figure}[h]
\includegraphics[width=\textwidth]{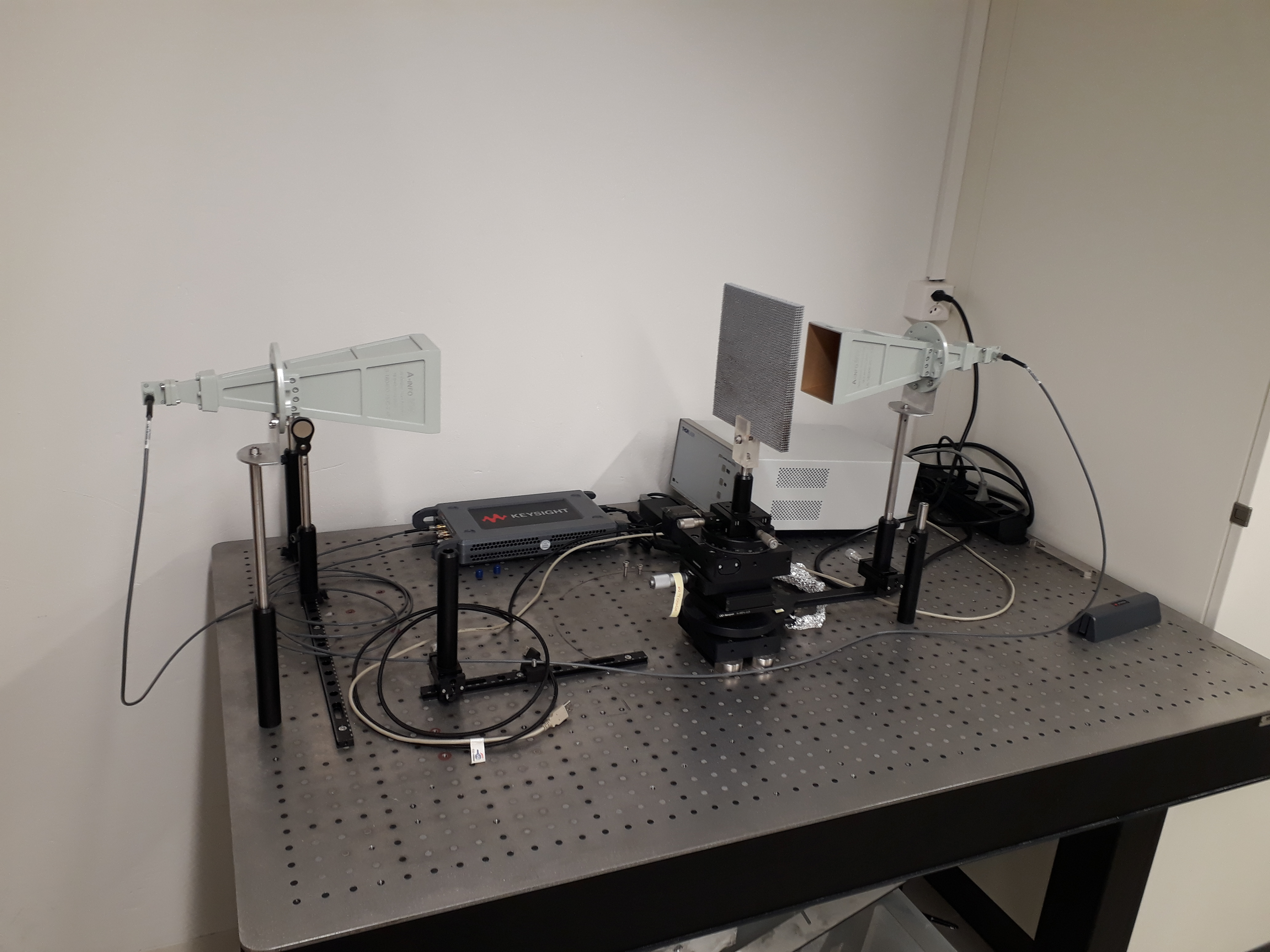}
\centering
\caption{Picture of the full setup without the microwave absorbing foam. The sample is placed on a computer-controlled rotating stage. The left antenna is used as a transmitter, and the right one as a receiver. Both antennas, LB-180400-25-C-2.4F (A-INFO Inc.), have identical specifications and were made by the same manufacturer.}
\label{fig:supp:microwave}
\end{figure}

\printbibliography